\documentclass[useAMS,usenatbib,usedcolumn]{mn2e} 
\usepackage{aasmacros}
\usepackage{amssymb,amsmath} 
\usepackage[pdftex]{graphicx} 
\usepackage[T1]{fontenc}
\usepackage{aecompl} 
\usepackage{url}
\usepackage{times} 
\usepackage{pdfsync} 
\usepackage[usenames,dvipsnames,svgnames,table]{xcolor}

\newcommand{\Sersic}{S\`{e}rsic}
\newcommand{\Devauc}{De Vaucouleurs}
\newcommand{\Msol}{\,\mathrm{M_{\sun}}}
\newcommand{\kpc}{\,\mathrm{kpc}}
\newcommand{\sbunits}{\,\mathrm{mag \, arcsec^{-2}}}
\newcommand{\sdunits}{\,\mathrm{M_{\sun} \, kpc^{-2}}}

\newcommand{\subfind}{{\sc subfind}}

\newcommand{\CA}[1]{#1}
\newcommand{\CB}[1]{#1}
\newcommand{\CC}[1]{#1}
\newcommand{\CD}[1]{#1}
\newcommand{\CE}[1]{#1}
\newcommand{\CX}[1]{#1}

\newcommand{\ssp}{SSP}

\DeclareRobustCommand{\Mass}[1]{M_{\mathrm{#1}}}
\DeclareRobustCommand{\Per}[1]{\%_{\mathrm{#1}}}
\DeclareRobustCommand{\SersicA}{\log_{10}\,\Sigma_{50}}
\DeclareRobustCommand{\SersicR}{R_{50}}
\DeclareRobustCommand{\Nprog}{N_{\mathrm{prog}}}
\DeclareRobustCommand{\Nf}{N_{\mathrm{50}}}
\DeclareRobustCommand{\Nn}{N_{\mathrm{90}}}

\title[BCG/ICL surface photometry in $\Lambda$CDM]{Surface photometry
of brightest cluster galaxies and intracluster stars in $\Lambda$CDM}

\author[Cooper et al.]{A. P. Cooper$^{1}$\thanks{E-mail:
acooper@nao.cas.cn},  L. Gao$^{1, 2}$, Q. Guo$^{1}$, C.S. Frenk$^{2}$, A. Jenkins$^{2}$, V. Springel$^{3,4}$
\newauthor and S.D.M. White$^{5}$\vspace{0.2cm} \\
$^{1}$National Astronomical Observatories, Chinese Academy of Sciences, 20A
Datun Road, Chaoyang, Beijing 100012, China\\ 
$^{2}$Institute for Computational
Cosmology, Department of Physics, University of Durham, South Road, Durham, DH1
3LE, UK\\
$^3$Heidelberg Institute for Theoretical Studies, Schloss-Wolfsbrunnenweg 35,
D-69118 Heidelberg, Germany\\
$^4$Zentrum f\"{u}r Astronomie der Universit\"{a}t Heidelberg,
ARI, M\"{o}nchhofstr. 12-14, D-69120 Heidelberg, Germany\\
$^{5}$Max-Planck-Institut f\"{u}r Astrophysik, 
Karl-Schwarzschild-Str. 1, D-85748, Garching, Germany}

\begin{document}

\date{Accepted 2015 May 7. Received 2015 April 1; in original form 2014 July 14}

\pagerange{\pageref{firstpage}--\pageref{lastpage}} \pubyear{2015}

\maketitle

\label{firstpage}
\begin{abstract} 

  We simulate the phase-space distribution of stellar mass in nine massive
  $\Lambda$CDM galaxy clusters by applying the semi-analytic particle tagging
  method of Cooper et al. to the Phoenix suite of high-resolution $N$-body
  simulations \CA{($M_{200} \approx 7.5$--$33\times10^{14}\mathrm{M_{\sun}}$)}.
  The resulting surface brightness (SB) profiles of brightest cluster galaxies
  (BCGs) match well to observations.  On average, stars formed in galaxies
  accreted by the BCG account for $\gtrsim 90$~per cent of its total mass (the
  remainder is formed in situ). In circular BCG-centred apertures, the
  superposition of multiple debris clouds (each $\gtrsim 10$ per cent of the
  total BCG mass) from different progenitors can result in an extensive outer
  diffuse component, qualitatively similar to a `cD envelope'.  These clouds
  typically originate from tidal stripping at $z \lesssim 1$ and comprise both
  streams and the extended envelopes of other massive galaxies in the cluster.
  \CX{Stars at very low SB contribute a significant fraction} of the total
  cluster stellar mass budget: in the central 1~$\mathrm{Mpc}^{2}$ of a
  $z\sim0.15$ cluster imaged at SDSS-like resolution, our fiducial model
  predicts 80--95 per cent of stellar mass below a SB of $\mu_{V} \sim 26.5
  \sbunits$ is associated with accreted stars in the envelope of the BCG.  The
  ratio of BCG stellar mass \CX{(including this diffuse component)} to
  total cluster stellar mass is $\sim30$~per cent.

\end{abstract}


\section{Introduction}

Observations of diffuse intracluster light (ICL) have shown that many stars in
galaxy groups and clusters are `outside' the galaxies themselves
\citep{Zwicky51}. The stars responsible for ICL are thought to have been
tidally stripped from cluster galaxies \citep{Gallagher72, Ostriker75,
Richstone75, White76, Hausman78, Merritt84}. On this basis, $N$-body
simulations have shown that the gross structural and dynamical characteristics
of ICL and the central (`brightest') galaxies of clusters (BCGs) can be
reproduced within the context of the $\Lambda$ cold dark matter ($\Lambda$CDM)
cosmogony \CE{\citep{Moore96, Dubinski98, Napolitano03, Murante04, Willman04,
Diemand05, SommerLarsen05, Rudick06, Murante07, Ruszkowski09, Dolag10,
Puchwein10, Oser10, Laporte13b}}. 

The surface brightness profile is one of the most easily studied observables of
BCGs. Idealized $N$-body simulations have been used to predict how these profiles
evolve through successive generations of mergers between cluster galaxies
\CE{\citep[e.g][]{Naab09b, Hilz12, Hilz13, Laporte12, Laporte13b}}.
However, these simulations do not usually include a realistic treatment of in
situ star formation in galaxies, and this limits their ability to make
quantitative predictions for real BGCs.  Specifically, gas dissipation strongly
biases the phase space distribution of stars relative to that of dark matter (DM),
such that stars generally trace the deepest parts of DM potential
wells \citep[e.g.][]{Gao04_attractor, Diemand05}. In detail this bias depends
on when, where and in what quantity stars form within the evolving hierarchy of
DM structures \citep{Frenk85}. Only some of the more recent cluster
$N$-body simulations have addressed the strong observational constraints on these
factors \citep[e.g.][]{Laporte13b}.

The baryonic processes regulating galaxy formation -- gas cooling, star
formation, `feedback' from supernovae and active galactic nuclei (AGN) -- can
be included self-consistently in hydrodynamic simulations, at least in
principle.  In practice, most of these processes act on scales below the
resolution of current hydrodynamic solvers and thus have to be implemented with
ad hoc semi-analytic recipes. Long run times make it hard to compare different
implementations and test numerical convergence, leading to large variations
between the results of different groups \citep[e.g.][]{Aquila} and uncertain
agreement with the statistics of the present-day galaxy population.
\CE{Although these problems are now arguably under control for simulations of
individual galaxies, they remain relevant for simulations of galaxy formation
in clusters.}  For example,  cluster simulations in which supernovae are the
only source of feedback result in BCG stellar masses much higher than those
observed \citep[e.g.][]{Oser10, Rudick11}. The authors of several recent
hydrodynamic simulations have claimed that AGN feedback is required to produce
realistic BCG masses and ICL fractions \citep{Puchwein10, Martizzi12a}.
\CX{The implementation of this feedback in simulations remains highly
uncertain, as do its effects on the structural properties and mass scaling
relations of BCGs \citep{Ragone13}.}

Semi-analytic models have had more success in making quantitative predictions
that agree well with the statistical properties of large surveys, including
observed scaling relations for BCGs \CE{\citep{AragonSalamanca98, Conroy07,
DeLucia07, Purcell07, Bower08, Guo11}}. However, most semi-analytic models
represent galactic structure with one-dimensional axisymmetric `disc' and
`bulge' density profiles of predetermined form\footnote{Some have recently
introduced `intracluster light' components that grow in mass through tidal
stripping, without specifying how those stars are distributed \citep{Monaco06,
Henriques10, Guo11, Contini14}.}.  As galaxies merge, changes in the mass and
size of their structural components are approximated with simple energy
conservation laws based on the virial theorem and parametrized to match
idealized $N$-body simulations \citep{Cole:2000aa, Guo11}.  Average scaling
relations can be studied with these approximations.  However, predictions for
observations of massive merger remnants, including BCG and ICL surface
brightness profiles, need a more detailed and self-consistent description of
how the stars from each individual progenitor evolve in phase space. 

Here we bridge this particular gap between $N$-body and semi-analytic models by
using the $N$-body particle-tagging technique of \citet[][hereafter
C13]{Cooper10,Cooper13} in combination with the ab initio $\Lambda$CDM
semi-analytic model of \citet[][hereafter G11]{Guo11}. This technique allows us
to make quantitative predictions for the six-dimensional phase space
distribution of stars that form in semi-analytic galaxies, which can be
compared directly to observations. Using the G11 model ensures that the star
formation histories in our simulations are compatible with the statistics of
the present-day galaxy population. Our aim is to make $\Lambda$CDM-based
predictions for the stellar mass surface density profiles of accreted stars in
cluster BCGs and their associated ICL, and to help interpret recent
observational work on the fraction of stellar mass in clusters that belongs to
these components.

We proceed as follows. Our simulations are described in
Section~\ref{sec:simulations}. Images of the simulated clusters are shown in
Section~\ref{sec:images}. Section~\ref{sec:profiles} describes the surface
density profiles of BCGs and compares them to observations.
Section~\ref{sec:substructure} analyses the shape of these profiles in terms of
the contribution of individual accreted substructures.
Section~\ref{sec:iclfraction} discusses the relative fractions of stars in
different stellar components. We summarize and conclude in
Section~\ref{sec:conclusions}. Appendix~\ref{appendix:numerical} discusses
important numerical issues.

\section{Method}
\label{sec:simulations}

\begin{table*} 
  
  \caption{Properties of the Phoenix cluster simulations \citep{Gao12} and new
  results for their BCGs discussed in this paper. \CX{All results are at
  $z=0$}. The table is ordered by $M_{200}$. From left to right columns show:
  (1) the simulation label; (2) $M_{200}$ $[10^{14} \Msol]$; (3)
  $R_{200}$$[\mathrm{Mpc}]$ (4) $M_{500}$ $[10^{14} \Msol]$; (5) redshift of
  half-mass assembly for the cluster DM, $z_{\mathrm{h}}$; (6) the
  total \CX{BCG stellar mass bound to the main \subfind{} halo}
  \textit{including stars associated with galaxies below the resolution limit},
  $\Mass{BCG}^{\mathrm{all}}$ $[10^{12} \Msol]$; (7) the fraction of this mass
  formed in situ; (8) the fraction accreted; (9) the fraction associated with
  sub-resolution galaxies (see text); (10) the fraction associated with streams
  from surviving galaxies with resolved haloes; (11)
  $\Mass{sat}^{\mathrm{min}}$ $[10^{12} \Msol]$, the total stellar mass
  \CX{within the cluster FoF group} bound to resolved DM subhaloes
  (i.e.  not including stars associated with sub-resolution haloes) -- the
  stellar mass bound to subhaloes within $R_{200}$ is lower by a factor of
  $0.52^{+0.14}_{-0.17}$ (median of nine clusters $\pm$ range); (12) the number of
  significant progenitors of the accreted component of the BCG [see
  section~\ref{sec:contrib}] (13) the number of progenitors accounting for
  50~per cent of the accreted mass of the BCG; (14) the number accounting for
  90~per cent; (15) the effective radius of the BCG (mean of three orthogonal
  projections) including sub-resolution galaxy stars $[\mathrm{kpc}]$; (16) the
  effective radius when sub-resolution galaxy stars are excluded
  $[\mathrm{kpc}]$.}

  \label{tab:phoenixstars}


  \begin{tabular}{@{}llll lll lll lll l ll}
   
      Name
      & $M_{200} $ 
      & $R_{200} $ 
      & $M_{500} $ 
      & $z_{\mathrm{h}} $ 
    & $\Mass{BCG}^{\mathrm{all}}$ 
    & $\Per{ins}$        
    & $\Per{acc}$        
    & $\Per{sub res}$    
    & $\Per{streams}$    
    & $\Mass{sat}^{\mathrm{min}}$ 
    & $\Nprog$           
    & $\Nf$              
    & $\Nn$              
    & $R_{50}^{\mathrm{all}}$      
    & $R_{50}^{\mathrm{excl}}$  \\ 

    {\tiny [1]}   & {\tiny [2]}     &  \tiny [3]    & \tiny [4]     &\tiny [5]     & \tiny[6]    & \tiny[7]    & \tiny[8]    & \tiny[9]    & \tiny[10]    & \tiny[11]    & \tiny[12]    & \tiny[13]    & \tiny[14]    & \tiny[15]    &  \tiny[16] \\
    \hline 
    C   & 7.527 & 1.899 & 5.884 & 0.76  & 3.94 & 8.3  & 91.7 & 18.1 & 21.8  & 6.95  & 9.76  & 6     & 61    & 103   & 82.4         \\
    E   & 8.176 & 1.912 & 6.414 & 0.91  & 3.51 & 12.0 & 88.0 & 24.1 & 21.6  & 9.14  & 15.9  & 7     & 105   & 125   & 86.0         \\
    D   & 8.481 & 1.899 & 6.239 & 0.46  & 3.08 & 14.3 & 85.7 & 25.9 & 27.6  & 9.98  & 17.8  & 8     & 101   & 39.0  & 89.8         \\
    A   & 9.000 & 1.937 & 7.278 & 1.17  & 4.15 & 8.4  & 91.6 & 28.4 & 25.3  & 10.4  & 25.1  & 22    & 274   & 185   & 155          \\ 
    F   & 10.93 & 2.067 & 8.324 & 1.10  & 3.84 & 13.7 & 86.3 & 13.2 & 25.4  & 12.0  & 25.2  & 11    & 234   & 132   & 117          \\
    B   & 11.31 & 2.090 & 8.387 & 0.46  & 4.70 & 6.5  & 93.5 & 29.8 & 30.0  & 10.8  & 7.80  & 5     & 74    & 178   & 181          \\
    H   & 15.55 & 2.310 & 11.88 & 0.21  & 6.80 & 19.5 & 80.5 & 28.1 & 26.5  & 25.7  & 13.6  & 9     & 190   & 186   & 100          \\
    G   & 15.75 & 2.334 & 10.82 & 0.18  & 3.75 & 9.9  & 90.1 & 33.2 & 30.5  & 19.3  & 17.0  & 7     & 190   & 296   & 171          \\
    I   & 33.03 & 2.933 & 25.50 & 0.56  & 15.0 & 5.48 & 94.5 & 30.6 & 35.2  & 43.2  & 43.1  & 27    & 657   & 314   & 284          \\

    \hline
  \end{tabular}
\end{table*}

\subsection{Definitions of BCG and ICL}

Explicit definitions for terms like BCG and ICL are essential, because galaxies
do not have clear boundaries \citep{Zwicky51}.  We call the galaxy at the
centre of the potential well of the cluster `the BCG' even though that central
galaxy may not be the brightest in all filters or apertures.  We {\em do not}
\CX{make any a priori assumptions about the existence of a separate
intracluster stellar component distinct from the BCG}\footnote{By considering
the ICL to be part of the BCG, we effectively ensure that the central galaxy is
the most luminous.}.  \CA{Instead, we treat the BCG as a single entity
consisting of all stars which are not bound to any subhaloes in the cluster (as
identified by the \subfind{} algorithm; \citealt{Springel01}). In addition to
stars bound to the main cluster halo, this includes a small fraction of
entirely unbound stars within $R_{200}$ ($\sim 1$ per cent of the BCG stellar
mass).} As we demonstrate below, this definition of the BCG includes many stars
in low surface brightness regions far from the centre of the cluster.

Among BCG stars, we distinguish \textit{accreted stars}, which have been
stripped from galaxies other than the BCG, from \textit{in situ} stars, which
formed directly from the cluster cooling flow. In a simulation, the distinction
between in situ and accreted stars is technically simple and almost
unambiguous\footnote{The main ambiguity is that in identifying a unique `main
branch' of the halo merger tree, particularly at high redshift when equal mass
mergers are common.}. This contrasts with dynamical definitions of the ICL,
which separate stars bound to the `cluster potential' from those bound to (the
stellar component of) the BCG \citep{Dolag10,Puchwein10,Rudick11}. We will
argue that the accreted/in situ separation is physically meaningful because ICL
phenomena in massive clusters are driven by trends in the accreted component
with $M_{200}$. 

Given these definitions, we only use the term ICL in a very loose sense, to
refer to the observational phenomenon of light from low surface brightness
regions in galaxy clusters. Whenever we compare to observations, we do not
\CX{perform} any BCG/ICL or in situ/accreted separation \citep[following, for
example, ][]{Lin04}.

\subsection{Phoenix and Millennium II}

Phoenix is a suite of nine high resolution DM-only $N$-body simulations of very
massive $\Lambda$CDM galaxy clusters, resimulated using a `zoom' technique with
initial conditions drawn from the Millennium Simulation \citep{Gao12}. The
particle mass ranges from $m_{\mathrm{p}}=6.1\times10^{6}\Msol$ (Ph-E) to
$2.5\times10^{7}\Msol$ (Ph-I), such that there are approximately $130$ million
particles within $r_{200}$ in each cluster halo at $z=0$. \CX{The
Plummer-equivalent force softening scale, $\varepsilon$, is $438$~pc at $z=0$.
This scale is kept fixed in comoving coordinates throughout the simulation (for
example, the physical value of $\varepsilon$ at $z=1$ is $160$~pc).
Interparticle forces become exactly Newtonian at a comoving separation of
$2.8\varepsilon = 1.23$~kpc.} Bound DM haloes and subhaloes were identified by
applying the \subfind{} algorithm to Friends-of-Friends \citep[FoF;][]{Davis85}
groups.  Relevant properties of the cluster halo in each simulation are
summarized in Table~\ref{tab:phoenixstars}.  The application of the G11
semi-analytic galaxy formation model to Phoenix is described separately by Guo
et al. (in preparation). In certain figures we show a sample of less massive
clusters from C13 based on the Millennium II simulation
\citep{BoylanKolchin10}, which has comparable resolution
($m_{\mathrm{p}}=9.4\times10^{6}\Msol$).

Our simulations use a cosmology compatible with \textit{WMAP} 1
($\Omega_{\mathrm{m}}=0.25$, $\Omega_{\mathrm{\Lambda}}=0.75$, $n_{s}=1$,
$\sigma_{8}=0.9$) and we assume a Hubble parameter $h=0.73$ throughout (we
convert data from other authors accordingly). More recent limits from the
cosmic microwave background amount to a rescaling in the mass, size and
abundance of our clusters, but do not alter the trends in our galaxy formation
model significantly, beyond their effect on star formation histories through
DM halo collapse times \citep{Guo13_wmap}. Such changes are likely to
be smaller than the current observational uncertainties in BCG masses and sizes
\citep[e.g.][]{Bernardi13}.

\subsection{Particle Tagging}
\label{sec:tagging}

Our particle tagging technique is described in \citet{Cooper10} and its
application to massive galaxies in C13, to which we refer the reader for
further details. The fundamental idea is to use a weighted subset of dark
matter particles in a cosmological $N$-body simulation to approximate the
phase space evolution of stellar populations. Our method improves on similar
techniques that have been applied to BCG models in the past
\citep{White80_gradients, Napolitano03, Gao04_attractor, Diemand05, Laporte12,
Laporte13b}.  Our application is novel in several respects: the tagging is
coupled directly to the star formation predictions of the semi-analytic model,
which in turn is constrained to reproduce the $z=0$ stellar mass function;
\CB{it tags in situ stars at the time of their formation, rather than
composite galaxies at the time of their accretion}; and the tagging method is
constrained to reproduce the $z=0$ mass--size relation for galaxies dominated
by in situ stars, as described below. 

We first process the merger trees of the simulation with the G11 model and
identify each `single age' stellar population (\ssp) that forms between two
successive snapshots (a galaxy at any time is the sum of many \ssp s). For each
\ssp, we identify all the DM particles in the corresponding halo at
the later of the two bracketing snapshots, rank them by their binding energy
and select a fixed fraction, $f_{\mathrm{mb}}$, in rank order starting from the
most bound particle. \CD{This approximates the end result of the
dissipative collapse of gas prior to star formation, namely that newly formed
stars are more tightly bound than the bulk of the DM in the halo
\citep{WhiteRees78}.} Each of these tightly bound particles is tagged with an
equal fraction of the stellar mass in the \ssp.  A single DM particle
can be tagged with stars from several \ssp{}s that form at different snapshots.

In an NFW halo, our constant-$f_{\mathrm{mb}}$ tagging method selects a
subregion of the overall DM distribution function that corresponds to
a truncated exponential density profile when integrated over velocity space and
one dimension of configuration space (i.e. the projection of the galaxy along
the line of sight; see C13). In a smoothly growing halo such profiles transform
into pure exponentials over time as particles diffuse above the initial cut-off
energy corresponding to $f_{\mathrm{mb}}$. The exponential scale-lengths of
these profiles scale systematically (albeit weakly) with the value of
$f_{\mathrm{mb}}$. As discussed in C13, for a given $N$-body halo of known mass
and concentration, the scale-length \CD{obtained from particle tagging} can
be predicted using a spherically symmetric and isotropic approximation to the
NFW distribution function \citep[e.g.][]{Widrow00}. 

\begin{figure*}
\includegraphics[width=0.9\textwidth, trim=0.25cm 0.25cm 0.25cm 0.25cm, clip=True]{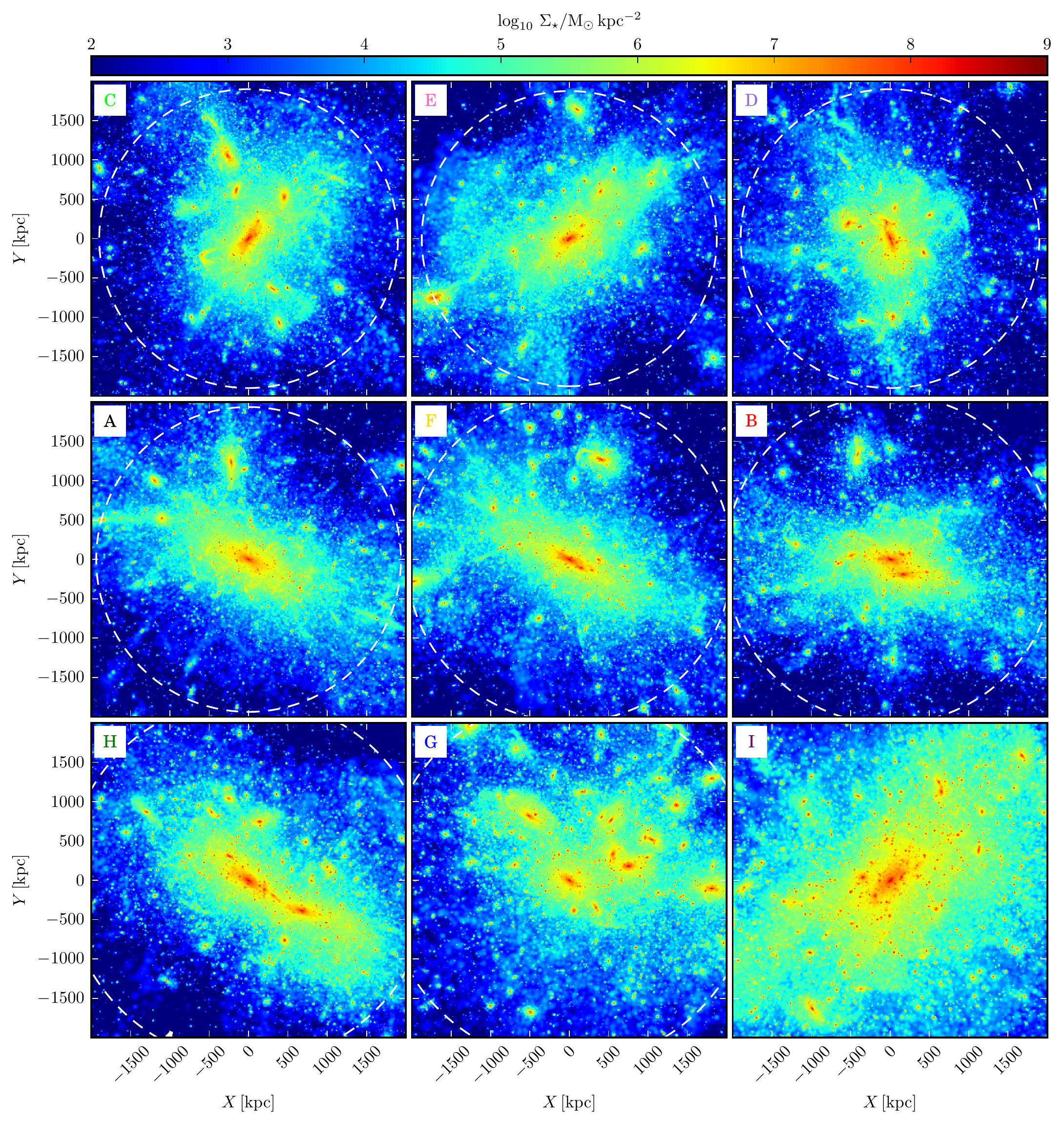}

\caption{Projected $3 \,  \mathrm{Mpc} \times3 \, \mathrm{Mpc}$ images of the
Phoenix clusters centred on their BCGs. $M_{200}$ increases from left to right
and top to bottom. The white dashed line shows $R_{200}$ (outside the image for
Ph-I).  The viewing angle is chosen randomly. Colours correspond to stellar
mass surface density on a $\log_{10}$ scale. Particles are smoothed by a cubic
spline kernel scaled by the density of their 64 nearest neighbours. `Hot spots'
are individual cluster galaxies; only very small scale density fluctuations are
due to shot noise. The brightest galaxies are surrounded by extensive diffuse
envelopes of tidal debris.} 

\label{fig:gallery} 

\end{figure*}

\CE{By assuming a constant universal value of $f_{\mathrm{mb}}$ and that all
stars in a given $z=0$ halo form at a single time}, \CD{it is
straightforward to apply this approximation to the population of DM
haloes in a cosmological simulation and thereby compute a} relation between
$M_{\star}$ and half-mass radius ($R_{50}$) for galaxies of $M_{\star} \lesssim
10^{10} \, \mathrm{M_{\sun}}$ (i.e. the regime where $M_{\star}$ is dominated
by in situ star formation rather than accretion; \citealt{Guo08}). \CD{The
size--mass relation inferred in this way will simply reflect the statistics of
the halo mass--concentration relation and the $M_{\star}$--$M_{200}$ relation,
but is nevertheless a good approximation to the results of full particle
tagging based on semi-analytic star formation histories}.  Specifically, once
the $M_{\star}$--$M_{200}$ relation is fixed, the normalization of
$R_{50}(M_{\star})$ in this approximation is determined by $f_{\mathrm{mb}}$.
By comparing such predictions to observations, C13 found a range of acceptable
values $1 \lesssim f_{\mathrm{mb}} \lesssim 3$ per cent. We use only
$f_{\mathrm{mb}} = 1$ per cent here (see Appendix~\ref{appendix:numerical}).   
 
As discussed in C13, \CB{our method may underestimate the central densities
and internal binding energies of late-type galaxies (and thus overestimate
their sizes)} because we \CE{do not explicitly model the dynamics of} the
dissipative collapse of cold gas and we neglect the adiabatic contraction of
DM due to baryons \citep[e.g.][]{Gnedin04}.  However, recent work
predicts only mild contraction for star formation efficiencies compatible with
observations \citep[][]{Dutton07, Abadi10, Schaller14a} and
also a counteracting expansion due to feedback \citep{Navarro96, Pontzen12}.
Moreover, this paper is focused on early-type galaxies, which result from
mergers occurring at $z<1$, after the majority of stars have formed in their
progenitors (in our model, in situ star formation triggered by low redshift
mergers only accounts for a small fraction of the mass in present-day BCGs).
Violent relaxation in low mass ratio mergers and dynamical friction acting on
infalling substructures can reduce the central density cusps created by any
dissipative contraction at high redshift and can simultaneously increase the
central DM mass fraction \citep{ElZant01, Gao04_attractor,
Ruszkowski09, Hilz12, Laporte12, Remus13}.  Strong lensing observations of the
total mass profiles of massive early-type galaxies imply very little nett
modification of the DM \CD{in the inner regions} despite the
central concentration of stars \citep{Newman13_totaldensity,
Newman13_separated, Dutton14}. \CE{On the other hand, hydrodynamical
simulations have shown that baryonic effects can alter the DM
distribution even in the outer parts of massive haloes \citep{vanDaalen11}}.

\section{Images}
\label{sec:images}

\begin{figure*}
\includegraphics[width=0.9\textwidth, trim=0.25cm 0.25cm 0.25cm 0.25cm, clip=True]{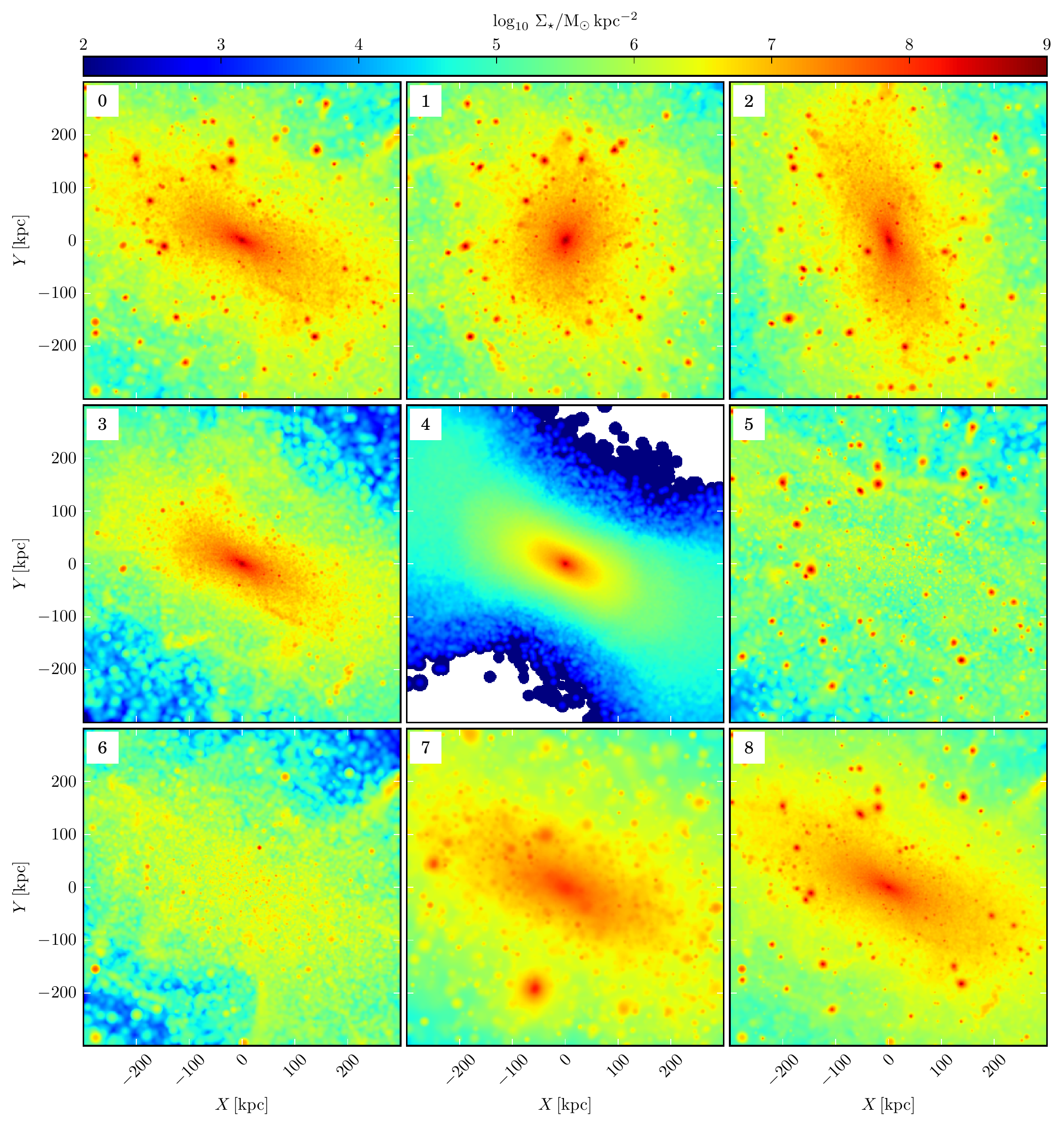}

\caption{A zoom-in to the \CX{central $600 \times 600$~kpc} region of Ph-A,
showing different subsets of particles in each panel. [0, 1, 2]: orthogonal
projections of all stars (note triaxiality); [3] accreted BCG stars
\CX{not associated with bound satellite galaxies or sub-resolution
galaxies in the semi-analytic model} (note absence of satellites, point-like
overdensities are massive stellar particles); \CE{[4] in situ BCG stars only
(note extent and elongation, due to halo response during last major merger);
[5]  all stars associated with surviving progenitor subhaloes (bound and
unbound; note coherent streams, \CX{mostly seen as linear yellow overdensities
around substructures}); [6] stars associated with `sub-resolution' haloes that
survive in the semi-analytic part of our model but not in the $N$-body simulation
(by default in our model, all these stars are assigned to the BCG; \CX{note
that the distribution of stars in this panel is somewhat more centrally
concentrated that those in panel 5, and more clearly aligned with the major
axis of the BCG}); } [7] all stars at Phoenix level 4 resolution (see
appendix~A; trivial timing and orbit differences mean satellite positions are
not identical; note convergence in density of BCG light); [8] all stars in a
model with $f_{\mathrm{mb}}=0.05$ (note slightly more diffuse centre of BCG).} 

\label{fig:gallery_extra} 
\end{figure*}

Fig.~\ref{fig:gallery} shows \CX{$3$~Mpc $\times 3$~Mpc} stellar mass surface
density images of the Phoenix clusters, including both stars bound to the
cluster potential (which we identify with the BCG) and stars bound to subhaloes
(which we identify with other cluster members). The `diffuse' light around the
BCG \CB{is very anisotropic, extends} to the edge of these images and contains
several relatively bright and coherent streams in most of our clusters (e.g. at
approximately $(X,Y) = (600, -250)$ in Ph-C). Extensive diffuse light and a
handful of bright streams have been observed in nearby clusters
\citep{Conselice99, Feldmeier04}, particularly Virgo \citep{Mihos05,
Janowiecki10, Rudick10}; Centaurus, with a $\sim170$~kpc $\times 3$~kpc stream
of $\mu_{R}\sim26.1 \, \mathrm{mag \, arcsec^{-2}}$ \citep{Roldan00}; and Coma
\citep[e.g.][]{Melnick77, Thuan77, Trentham98, Adami05b} which has a broad
stream of length $\sim130$~kpc, width $15$--$30$~kpc, $ \mu_{R}\sim25.7 \,
\mathrm{mag \, arcsec^{-2}}$ and no obvious progenitor \citep{Gregg98}.

\begin{figure*}
\includegraphics[width=168mm, trim=0.1cm 0cm 0.1cm 0cm, clip=True]{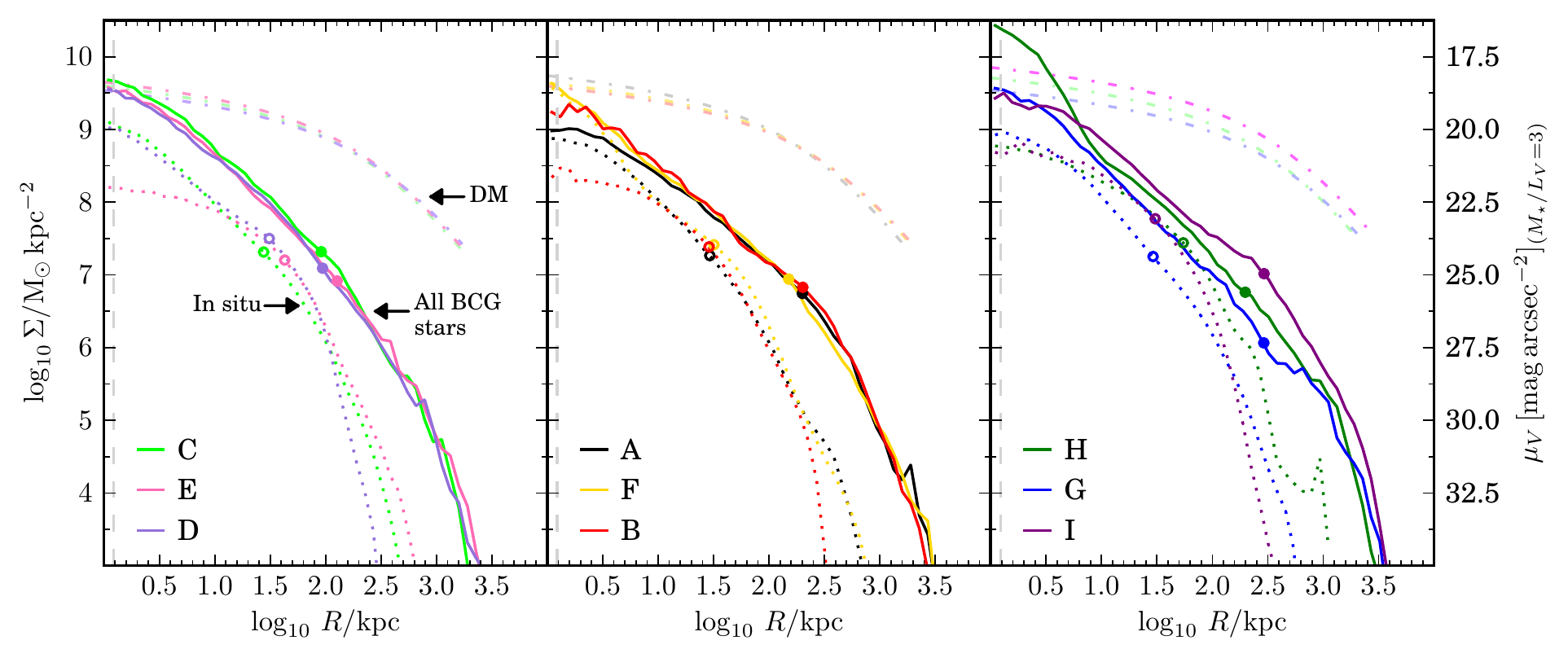}

\caption{\CA{Surface density profiles of BCG stars in Phoenix clusters A--I
(solid lines; \CD{simulations} are ordered by $M_{200}$).  Dotted lines show
the profiles for in situ stars only and dot--dashed lines plot the surface
density of the best-fit NFW DM profile out to the virial radius \citep{Gao12}}.
Filled and open circles mark the half-mass radii of the total and in situ
profiles respectively ($R_{50}$).  A vertical grey dashed line indicates the
simulation softening length. The right-hand axis gives an approximate
conversion of $\Sigma_{\star}$ to $V$-band surface brightness, assuming
$M_{\star}/L_{V}=2.5$.  Clusters of similar mass have similar profiles, with
accreted stars dominating at almost all radii.} 

\label{fig:basic_density_profiles} 
\end{figure*}

Fig.~\ref{fig:gallery_extra} shows the \CX{central $600 \times 600$~kpc}
region of Ph-A in more detail. Panels 0, 1 and 2 show how the appearance of the
BCG changes with projection along orthogonal axes of the simulation box -- all
BCGs are significantly elongated along one axis, although this has little
effect on any of the properties described in this paper. Panel~3 shows only
stars that have been accreted by the BCG and which are not associated with
\textit{any} surviving satellite galaxy in the semi-analytic model
\CX{(including sub-resolution galaxies)}; panel~4 shows stars formed in situ
in the main branch of the BCG's merger tree. These components show similar
alignment and elongation\footnote{\CC{Our simulation does not include baryonic
effects which could make the potential more spherical, so this elongation may
be exaggerated.}}, both of which can be attributed to the one or two most
recent `major' merger events (see e.g.  \citealt{Cooper11b} for a high
resolution visualization of this effect in a less massive elliptical galaxy). 

Panel~5 of Fig.~\ref{fig:gallery_extra} shows stars associated with surviving
well-resolved galaxies, including those that are still bound (visible as
obvious concentrations) and those that have been stripped into the accreted
component of the BCG. Many of the bright galaxies around the BCG contribute a
substantial mass of tidal debris, to the extent that the overall distribution
looks rather uniform because individual tidal streams have low contrast against
the bulk of diffuse material.

\subsection{Stars associated with sub-resolution haloes and other numerical
issues}

In the G11 model, semi-analytic galaxies can survive even after tidal stripping
reduces the mass of their associated DM subhalo below the 20-particle
resolution limit of \subfind. G11 assume that no stars are stripped from these
galaxies associated with unresolved haloes (called orphans or `type 2' galaxies
elsewhere) until they merge or are (instantaneously) destroyed by tides. In
this paper, we choose to assign all of the stars associated with these galaxies
to the accreted component of the BCG. In other words, we assume that subhaloes
will have lost essentially all of their stellar mass by the time they are
stripped below a total mass of 20 particles ($\sim2\times10^{8} \Msol$).

\CE{A detailed discussion of this variation to the G11 model is given in
Appendix~\ref{appendix:numerical}, where we argue it is justified by the
neglect of stellar stripping in G11}. To summarize, in the case of a Milky
Way-like halo having $M_{200}\sim10^{12}\Msol$ at infall, the 20 particle
resolution limit corresponds to a remnant halo \CA{mass fraction of only
$\sim0.02$ per cent}; \CD{it seems implausible that $100$ per cent of the
stars in such a halo would remain bound after such dramatic DM mass
loss, as the G11 model assumes}.  \CD{However, for dwarf galaxy haloes with
maximum pre-infall mass close to $10^{8} \Msol$, the corresponding fractional
mass-loss required to fall below the resolution limit is much smaller and hence
the assumption that \textit{all} stars bound to unresolved halo remnants have
been stripped is less accurate}. The nett contribution of stars from these
\CD{less massive} haloes could, \CD{in principle,} make a significant
difference to our conclusions.  \CB{In Appendix~\ref{appendix:numerical} we
show alternative results for the case where all stars from sub-resolution
haloes are excluded from the definition of the BCG (and hence treated as stars
in surviving galaxies)}.

\CX{Panel~6} of Fig.~\ref{fig:gallery_extra} shows the subset of stars in
question, those that belong to the accreted component of the BCG according to
the $N$-body part of our model but are associated with surviving sub-resolution
haloes in the semi analytic part. Clearly (almost by definition) the vast
majority of these are not concentrated in galaxy-like clumps. Rather, they are
spread out all over the halo, much like the stars \CX{being stripped} from
well resolved subhaloes (\CX{panel 5}).  They are more centrally concentrated
in the cluster, consistent with the defunct subhaloes either having radial
orbits (those lost due to rapid disruption) or having orbited for a long time
in the centre of the cluster (those lost due to prolonged stripping).

Panel~7 of Fig.~\ref{fig:gallery_extra} shows the same halo simulated at lower
resolution and panel~8 the result of choosing $f_{\mathrm{mb}}=5$ per cent at
our standard resolution -- neither of these makes a substantial difference to
the overall appearance of the BCG, although there are subtle changes to the
density distribution. Appendix~\ref{appendix:numerical} also addresses these
numerical issues in more detail.

\section{Surface density/brightness profiles}
\label{sec:profiles}

\subsection{Phoenix clusters}

Fig.~\ref{fig:basic_density_profiles} shows $\Sigma_{\star}(R)$, the
azimuthally averaged\footnote{Throughout this paper we only consider profiles
in circular apertures. Both \citet{Zibetti05} and \citet{Seigar07} find that
changes in the profile of BCG ellipticity and position angle do not correspond
directly to inflections in the surface brightness profile.} stellar mass
surface density of stars associated with the BCG in each Phoenix cluster at
$z=0$, according to the definition in Section~1, i.e. all stars tagged to DM
particles bound to the potential well of the cluster's DM halo
\CA{but} not to any of its subhaloes. 

Although our clusters span a factor of $4$ in $M_{200}$, their central galaxies
have remarkably similar $\Sigma_{\star}(R)$ profiles.  The cluster-to-cluster
range in surface density across our sample is $\sim 0.5$ dex at almost all
radii (and considerably less if Ph-I is excluded).  Stochastic variations in
individual profiles obscure any trend of profile shape with $M_{200}$,
particularly for our most massive haloes; there is a weak trend of amplitude
consistent with the expected correlation between $M_{\star}$ and $M_{200}$ (see
Table \ref{tab:phoenixstars}). The central surface density ($R<10$~kpc) is
notably lower in Ph-A (the oldest cluster) and an order of magnitude higher in
Ph-H (the second youngest, which has a complex core structure).  These
differences hold for different random choices of projection, because projection
effects are generally smaller than the scatter between our clusters.  The
largest differences between clusters and between different projections for any
given cluster are at $R<10$~kpc, where neglecting the gravity of baryons makes
our model less reliable in any case. 

Accreted stars dominate over in situ stars (dotted lines) at almost all radii,
including the most luminous central regions of the BCG -- the notable
exceptions are Ph-A and Ph-F (the two oldest clusters) in which the ratio is
almost $1:1$ within $R<5$~kpc. This dominance of accreted stars at all radii
distinguishes BCGs in very massive clusters from those in haloes of
$M_{200}\lesssim10^{14} \mathrm{M_{\sun}}$, where in situ stars typically
dominate within $\sim10$ kpc at $z=0$ (see C13) and influence the total surface
brightness profile significantly out to $\sim30$~kpc. It is also notable that
the profiles of in situ stars in Fig.~\ref{fig:basic_density_profiles}
typically have a similar shape to the total light profile \CB{-- C13 found
	that this is not the case in less massive haloes}. This indicates
	in situ stars and accreted stars are \CB{relatively} well-mixed in
	the BCGs of massive clusters.

Our BCGs are an order of magnitude more massive than those studied at
comparable resolution by C13.  Their half-mass radii ($R_{50}$; see
Table~\ref{tab:phoenixprofiles}) nevertheless lie on an extrapolation of the
trend shown in C13 for early-type galaxies above $\sim10^{11.3} \Msol$,
suggesting no further steepening of this relation in the regime where size
growth is dominated by accretion.  This behaviour agrees roughly with the
observed relations of \citet{Guo09} and \citet{Bernardi12_arxiv}, although our
BCGs are $\sim0.2$~dex larger at a fixed stellar mass compared to an
extrapolation of the SerExp relation preferred by
\citet{Bernardi12_arxiv}. \CD{At the resolution of Phoenix,}
excluding stars from sub-resolution haloes does not affect this result
significantly.

\subsection{Comparison to observations}

\begin{figure}
\includegraphics[width=84mm, trim=0.1cm 0cm 0.1cm 0cm, clip=True]{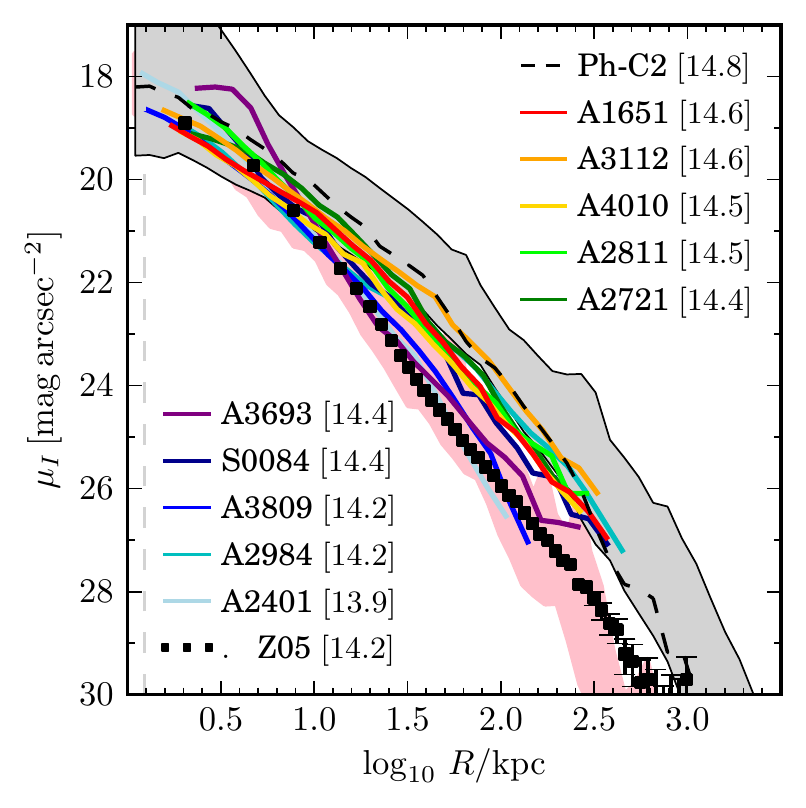}

\caption{Envelope of simulated surface brightness profiles in Phoenix (grey
	region) and Millennium II ($13.8<\log_{10}M_{500}/\Msol<14.0$, pink
	region) compared to observations in the Cousins $I$ band from
	\citet[][+0.429 mag to convert to AB, corrected for $(1+z)^4$ SB
	dimming]{Gonzalez05}. The black dashed line shows one random projection
	of the least massive Phoenix cluster, Ph-C, for reference.  The legend
	indicates Abell catalogue (A: north, S: south) \CB{and, in
		parenthesis, the MCXC $\log_{10} \, M_{500}$ value} from
		\citet{Piffaretti11}.  Squares with error bars show the stacked
		BCG profile of \citet{Zibetti05} in the SDSS $i$ band (assuming
		$i-I = 0.1$~mag). The shape and amplitude of simulated profiles
		and their trend with $M_{500}$ agree well with these data.} 

\label{fig:iband} \end{figure} We now compare the amplitude and shape of our
simulated BCG surface brightness profiles to observations.  There are many deep
BCG surface brightness profiles in the literature, covering a wide range of
galaxy and halo masses \citep{Schombert86, Schombert88,  Mackie90, Uson91,
Graham96, Gonzalez05, Krick06,  Patel06, Krick07, Seigar07, Bildfell08,
Donzelli11}. However, the majority of these are likely to correspond to
clusters much less massive than those in the Phoenix sample. C13 found that the
shape and amplitude of surface brightness profiles simulated with our technique
are strongly correlated with $M_{200}$. A shallow $M_{\star}$--$M_{200}$
relation means that comparison at fixed $M_{\star}$ introduces considerable
scatter to these trends. Therefore, we choose observed clusters for comparison
according to estimates of the total mass enclosed within particular contours of
overdensity (e.g. 500 times the critical density of the universe, denoted
$M_{500}$).

The precision of most cluster mass measurements is likely to be no better than
$\sim25$ per cent and the absolute calibration can vary by even more between
different studies (see \citealt[][]{Rozo14_scalings} and \citealt{Applegate14}
for recent discussions). Nevertheless, they become increasingly accurate for
massive clusters where a variety of estimators can be applied, to the point
where they are likely more reliable than surface brightness limited $M_{\star}$
measurements for BCGs. Moreover, cluster mass estimates are usually independent
of the BCG surface photometry we wish to compare with (this is not
the case for stellar masses; e.g.  \citealt{Bernardi13}). We obtain aperture
mass measurements from the MCXC catalogue of \citet{Piffaretti11}, \CD{who
standardized heterogeneous X-ray luminosity data to create a single catalogue
of self-consistent $M_{500}$ estimates}.

\begin{figure}
\includegraphics[width=84mm, trim=0.1cm 0cm 0.1cm 0cm, clip=True]{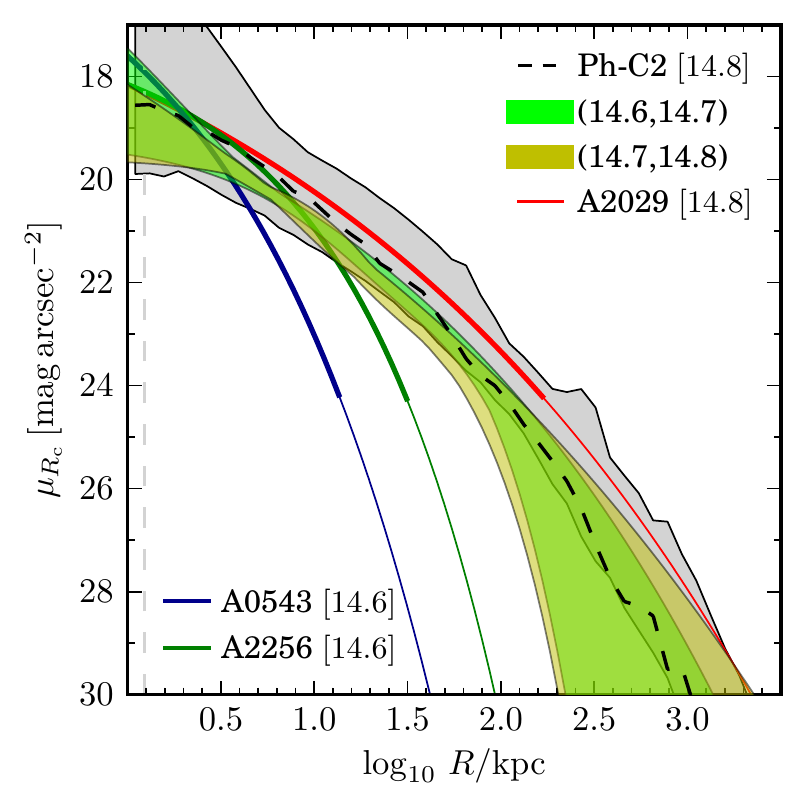}

\caption{Comparison to observations in the $R_{\mathrm{c}}$ band from
\citet[][converted from Vega to AB by adding 0.117 mag and corrected for SB
dimming]{Donzelli11}. Shaded areas (green and yellow) are envelopes of
best-fitting profiles for BCGs with MCXC masses $14.6 < \log_{10} \, M_{500} /
\Msol < 14.7$ (four galaxies, green) and $14.7 < \log_{10} \, M_{500} / \Msol <
14.8$ (seven galaxies, yellow). Two galaxies in the lower interval, (A0534 and
A2256), have extremely concentrated profiles; we show these individually, with
thinner lines where the fits are extrapolated. Abell 2029 (red line) is the
only galaxy in \citet{Donzelli11} with an MCXC mass $\log_{10} \, M_{500} /
\Msol > 14.8$.  The Phoenix simulations also agree well with these data, except
for the two outliers (compare Fig.~\ref{fig:iband}).}

\label{fig:Rcband} 
\end{figure}

\begin{figure}
\includegraphics[width=84mm, trim=0.1cm 0cm 0.1cm 0cm, clip=True]{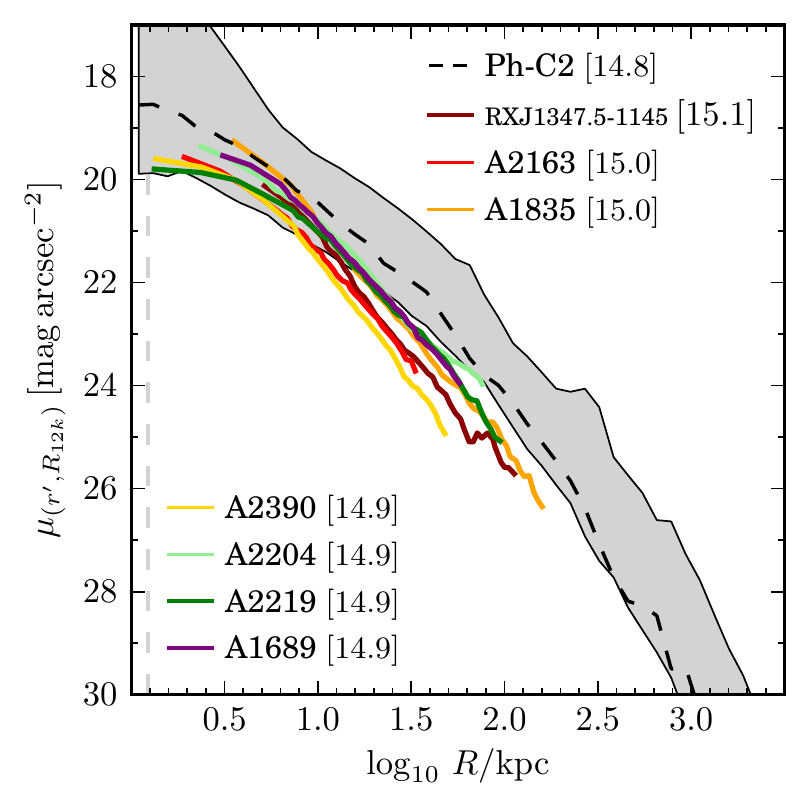}

\caption{Comparison to observations in the MegaCam $r^{\prime}$ and CFHT 12K
camera Mould $R$ bands from \citet[][converted from Vega to AB by adding 0.138
mag, and corrected for SB dimming; we assume no colour offset between these two
filters]{Bildfell08}. Galaxies were selected to have MCXC masses $
\log_{10}\,{M_{500}/\Msol > 14.9}$. The simulations do not match these data as
well as those in previous figures; this could point to systematic errors in the
photometry and/or $M_{500}$, or else to an overproduction of BCG stellar mass
in our model (compare Fig.~\ref{fig:iband}).} 

\label{fig:bildfell} 
\end{figure}

In Fig.~\ref{fig:iband} we show $I$-band surface brightness\footnote{\CX{In
Figs.~\ref{fig:iband} -- \ref{fig:bildfell}, surface brightness is obtained by
applying the \citet{Bruzual:2003aa} population synthesis model to the star formation
history of each tagged particle,  assuming a universal \citet{Chabrier03} IMF
and instantaneous recycling with parameters given in G11. The resulting
spectral energy distribution is then convolved with an appropriate transmission
curve to determine the mass-to-light ratio of the particle in the relevant
band.}} profiles from the catalogue of \citet{Gonzalez05}.  Matching against
MCXC yields 10 BCGs in common, of which three have $\log_{10}\,{M_{500}/\Msol >
14.5}$.  Our least massive Phoenix halo is Ph-C (profile shown by a black
dashed line) which has $\log_{10}\,{M_{500}/\Msol=14.9}$. The envelope of the
Phoenix profiles is indicated by a grey shaded region. The \citet{Gonzalez05}
profiles have similar amplitude to one another at $10 \kpc$, $\sim 1 \sbunits$
below the mean of our simulations but within the lower envelope. The data have
a weak trend towards steeper slopes at lower $M_{500}$, such that only the most
massive (e.g.  Abell 1651, Abell 3112) have shapes in good agreement with the
simulations. The steeper slope and lower amplitude of profiles from less
massive BCGs is, however, in good agreement with the 16 clusters of mass  $
13.8 < \log_{10}\,{M_{500}/\Msol < 14.0}$ simulated with the same technique by
C13 (pink shaded region). We conclude that our models are consistent with the
\citet{Gonzalez05} data\CX{; that these data lie below the median of the
simulations in Fig.~\ref{fig:iband} is simply because the Phoenix haloes have
systematically higher $M_{500}$}.

Fig~\ref{fig:iband} also shows data from \citet{Zibetti05}, who stacked SDSS
$i$-band images of $z\sim0.25$ BCGs to derive an average surface brightness
profile. These data are in agreement with the individual profiles of
\citet{Gonzalez05} for masses $\log_{10}\,{M_{500}/\Msol \lesssim 14.2}$ and
with our Millennium II results. As noted in C13, this is consistent with
estimates of the mean halo mass of the \citet{Zibetti05} sample based on
richness \citep{Rozo09}.

Fig.~\ref{fig:Rcband} presents a similar comparison to the Cousins $R$-band
data of \citet[][]{Donzelli11}, published as either one- or two-component fits
to regions $\mu_{R_{c}} > 24.5 \sbunits$. Yellow and green shaded regions show
the envelope of best-fit profiles for BCGs in this sample that can be matched
to the MCXC catalogue in the mass ranges $ 14.6 < \log_{10}\,{M_{500}/\Msol <
14.7}$ (four galaxies) and $ 14.7 < \log_{10}\,{M_{500}/\Msol < 14.8}$ (seven
galaxies), respectively. These agree well with the Phoenix haloes in the range of
$\mu_{R_{c}}$ used for the fit. Three galaxies from \citet[][]{Donzelli11} are
plotted individually. Abell 2029 (red line) is the only cluster in
\citeauthor{Donzelli11} matched to an MCXC cluster with
$\log_{10}\,{M_{500}/\Msol > 14.8}$. This most massive BCG agrees particularly
well with our simulations.  The fits for Abell 543 and Abell 2256 are very
different to the other haloes in their $M_{500}$ range, suggesting either that
those clusters are atypical, that there are issues with their photometry in
\citet[][]{Donzelli11}, or that their MCXC halo masses are greatly
overestimated (by more than an order of magnitude according to the predicted
trends of C13).

Finally, Fig.~\ref{fig:bildfell} compares our simulations with the data of
\citet{Bildfell08} in either the CFHT/MegaCam $r'$-band or the CFHT/12K
$R$-band.  We select MCXC-matched clusters with $ \log_{10}\,{M_{500}/\Msol >
14.9}$.  This sample should be well-suited to comparison with our simulations
because several of the most massive clusters in the MCXC catalogue are
included.  However, given the good agreement with less massive clusters seen in
previous figures, the discrepancies in Fig.~\ref{fig:bildfell} are surprising.
The simulations overpredict the observed profiles by 1--2 magnitudes at
$30\kpc$.  

Curiously, the approximate trend of amplitude with halo mass seen in the data
of \citet{Gonzalez07} is absent from the data of \citeauthor{Bildfell08} (and
remains so if less massive cluster profiles in their sample are included). If
significant errors in photometry\footnote{\textit{Hubble Space Telescope}
(\textit{HST}) photometry of Abell 2390 presented by
\citet{Newman13_totaldensity} agrees very well (in shape) with the relatively
steep profile of \citet{Bildfell08}.  \citeauthor{Newman13_totaldensity} also
obtain $M_{500} \sim 10^{15.1} \Msol$ based on weak and strong lensing, which
makes the disagreement with our model marginally worse \citep[see
also][]{Applegate14}. } and $M_{500}$ can be ruled out, the discrepancy in
Fig.~\ref{fig:bildfell}  may point to a systematic problem with our model --
for example, the luminosity of simulated BCGs may increase too rapidly with
$M_{200}$. This could also explain the apparent $\lesssim 0.2$ dex excess of
$R_{50}$ at $M_{200}\sim10^{14} \Msol$ in our simulations with respect to
observed relations (see previous section and C13).

\subsection{Functional forms}
\label{sec:functional_forms}
\begin{table*} 
  
  \caption{Fits to BCG stellar mass surface density profiles (listed in order
  of increasing $M_{200}$). Following the halo label, columns show groups
  ($R_{50}$ $\mathrm{[kpc]}$, $\log_{10} \Sigma_{50}/\mathrm{M_{\sun}\,
  kpc^{-2}}$, $n$) corresponding to the parameters of scale radius, surface
  density and \Sersic{} index in equation \ref{eq:sersicprof}. From left to
  right, these groups correspond to fits of a single \Devauc{} profile, a
  single \Sersic{} profile, and a two-component \Sersic{} profile. The latter
  is split into principle (1) and secondary (2) components according their
  contribution to the total mass integrated over the radial range $1 < R <
  10^{3.5}$~kpc. The final column gives $f_{2}$, the mass fraction of the
  secondary component. See text for details of the constraints on each profile
  model.}

  \label{tab:phoenixprofiles}

  \begin{tabular}{@{}l lcl lcl lcl lcl l}
   
    & 
    & $R^{1/4}$         
    & 
    & 
    & Single \Sersic{}       
    & 
    & 
    & Double \Sersic{} (1)   
    & 
    & 
    & Double \Sersic{} (2)   
    & 
    & \\ 

      Name
    & $\SersicR$  
    & $\SersicA$  
    & $n$         
    & $\SersicR$  
    & $\SersicA$  
    & $n$         
    & $\SersicR$  
    & $\SersicA$  
    & $n$         
    & $\SersicR$  
    & $\SersicA$  
    & $n$         
    & $f_{\mathrm{2}}$ \\ 
    \hline
  

C	&  87.2	& 7.28	& 4.00  &  107	& 7.15	& 4.09	    &  144	& 6.89	& 3.23  &  9.46	& 8.41	& 4.48 	& 0.14	\\ 
E	&  64.8	& 7.39	& 4.00  &  117	& 7.00	& 4.64	    &  60.0	& 7.38	& 4.96	&  377	& 5.77	& 1.66 	& 0.37	\\ 
D	&  135	& 6.85	& 4.00  &  128	& 6.95	& 3.29	    &  179	& 6.65	& 2.51  &  15.0	& 7.93	& 4.56 	& 0.15	\\ 
A	&  181	& 6.64	& 4.00  &  189	& 6.71	& 3.29	    &  215	& 6.63	& 2.77  &  6.86	& 8.24	& 4.81 	& 0.05  \\ 
F	&  148	& 6.85	& 4.00  &  156	& 6.84	& 3.87	    &  191	& 6.67	& 3.19  &  7.84	& 8.34	& 4.34 	& 0.08  \\ 
B	&  93.9	& 7.17	& 4.00  &  132	& 7.02	& 3.39	    &  263	& 6.49	& 0.96  &  75.6	& 7.14	& 6.53 	& 0.47	\\ 
H	&  169	& 6.89	& 4.00  &  196	& 6.81	& 3.74	    &  98.2	& 7.16	& 6.97  &  646	& 5.71	& 0.624 & 0.34	\\ 
G	&  48.3	& 7.42	& 4.00  &  309	& 6.04	& 9.07	    &  217	& 6.25	& 9.55  &  819	& 5.11	& 0.501 & 0.23	\\ 
I	&  179	& 7.04	& 4.00  &  279	& 6.87	& 3.35	    &  515	& 6.51	& 1.52  &  46.7	& 7.74	& 3.69 	& 0.18	\\ 

    \hline
  \end{tabular}
\end{table*}

\begin{figure}
\includegraphics[width=84mm, trim=0.1cm 0cm 0.1cm 0cm, clip=True]{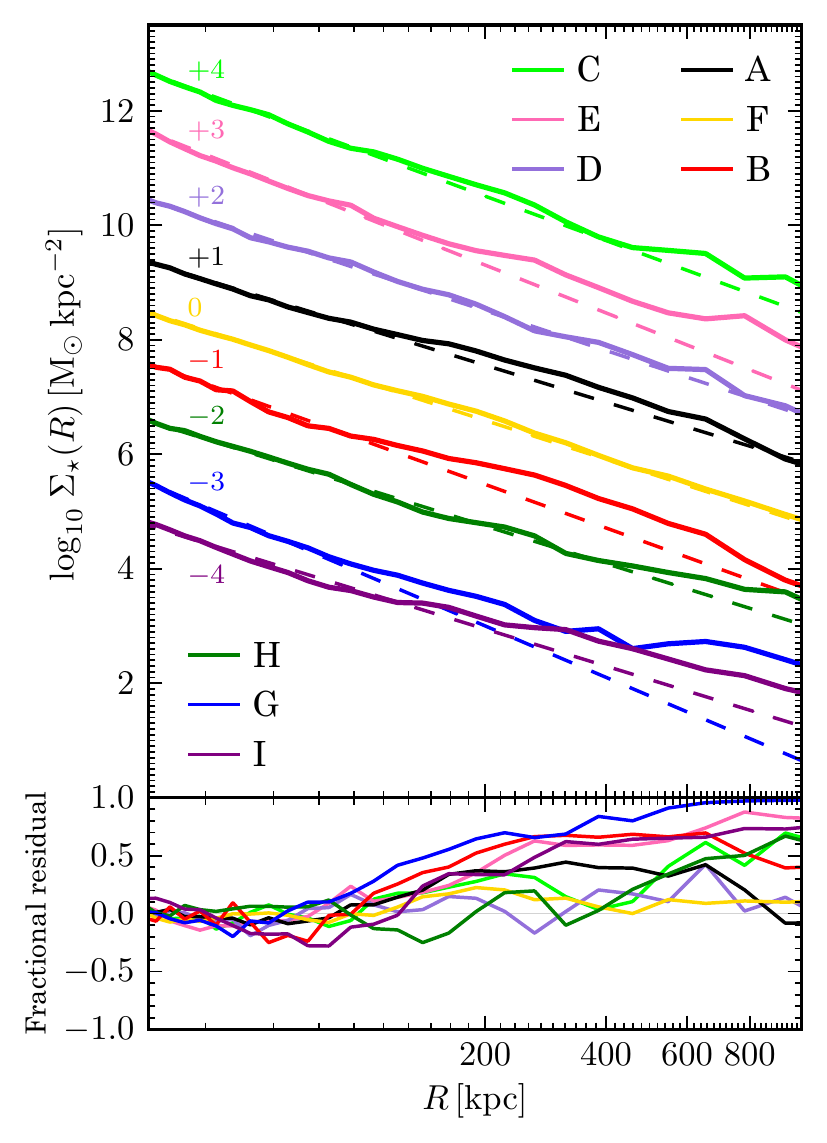}

\caption{Surface mass density of BCG+ICL stars, with the $x$-axis scaled such
that a $\Sigma_{\star} \propto R^{1/4}$ profile is a straight line. For
clarity, each profile is offset in $\log_{10} \Sigma_{\star}$ by $\pm 1$ to
$4$~dex as indicated ($M_{200}$ increases in order from top to bottom).  Dashed
lines correspond to $R^{1/4}$ fits to regions more dense than
$\Sigma_{\star}=10^{7}\sdunits$.  Fractional residuals are shown in the lower
panel. cD-like upturns relative to the inner $R^{1/4}$ fit occur at $R \gtrsim
200$~kpc in at least \CB{five clusters (A, B, E, G, I)}.}

\label{fig:devprof} 
\end{figure}

The so-called `cD envelope' phenomenon \citep{Matthews64,Oemler76} refers to an
`upturn' in surface density relative to the extrapolation of a standard profile
(often, but not strictly, an $R^{1/4}$ law) fit to the high surface brightness
regions of a BCG \citep[e.g.][]{Schombert88, Liu08}. The visual impression of
these `haloes' is particularly striking on deep photographic plates.
Fig.~\ref{fig:devprof} shows maximum likelihood $R^{1/4}$ profile fits to our
simulations in regions $\Sigma_{\star} > 10^{7}\sdunits$, corresponding to a
surface brightness $\mu_{R} \lesssim 26.5 \sbunits$ \citep[e.g.][]{Seigar07}
and radii roughly $R \lesssim 100 \, \mathrm{kpc}$.  Values of amplitude and
scale radius for each halo are given in Table~\ref{tab:phoenixprofiles}.
Upward deviations from these fits at larger radii are prominent in haloes B, E,
G and I, and also apparent in A, C and H. Halos D and F show very little
deviation from a single $R^{1/4}$ law. There is, at most, a very weak trend of
increasing excess with $M_{200}$ or halo assembly time.

\begin{figure} \includegraphics[width=84mm, trim=0.1cm 0cm 0.1cm 0cm,
  clip=True]{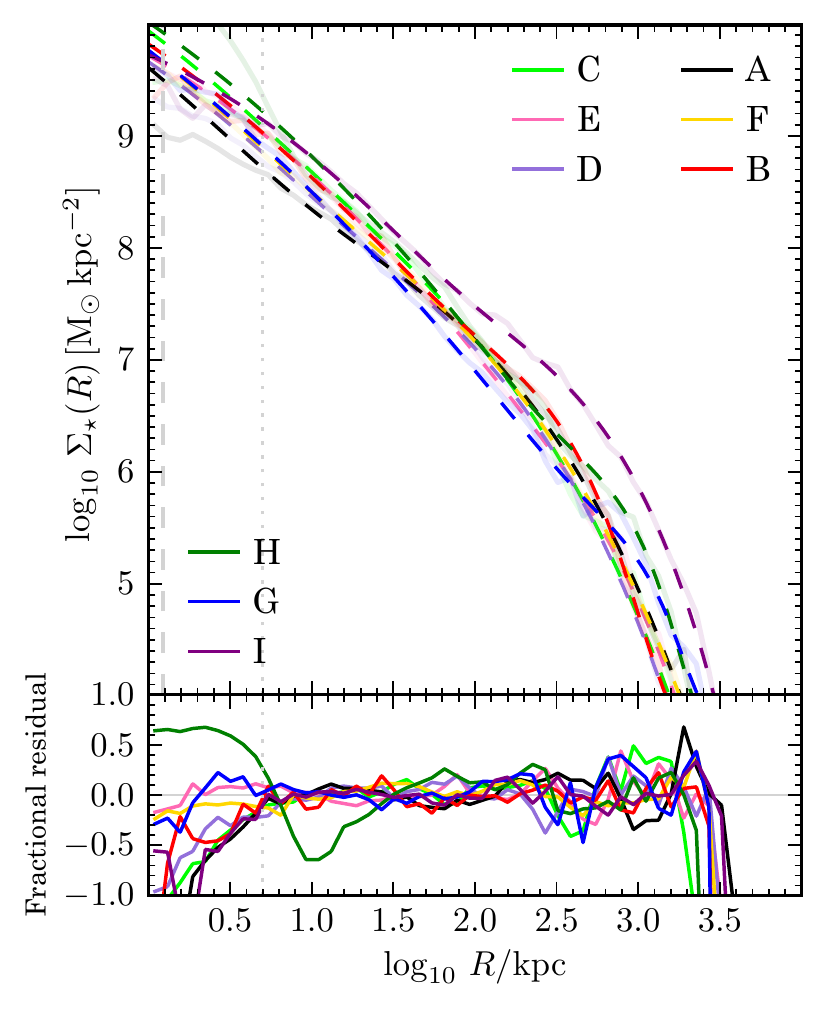}

\caption{Upper panel: dashed lines are double \Sersic{} profile fits to
	arbitrary projections of BCG+ICL stars over the range of radius and
	density shown in the figure. Simulation data are shown as faint solid
	lines of the same colour. Parameters are given in
	Table~\ref{tab:phoenixprofiles}. Lower panel: fractional residuals of
	each fit.  \CB{These maximum likelihood fits assume a Gaussian prior
	on $\bar{\Sigma_{\star}(R)}$ at each data point}, with a fiducial
	dispersion $\sigma=5\%$, and exclude the innermost $5$~kpc (dotted
	line).  Double \Sersic{} profiles fit well in the range $10$~kpc to
	$1$~Mpc; components with large half-mass radii tend to have $n\lesssim2$.}

\label{fig:sersicprof} 
\end{figure}

This `upturn' simply reflects the now well-known fact that massive BCGs are not
well fitted by $R^{1/4}$ profiles and does not reveal much about the physical
significance of the `excess' light  \citep{Lugger84,Seigar07,Schombert13}. The
more general \Sersic{} function \begin{equation} \Sigma(R) =
  \Sigma_{\mathrm{50}} \exp \{ -b_{n}[ \left( R/R_{\mathrm{50}} \right)^{1/n}
  -1 ]\} \label{eq:sersicprof} \end{equation} provides a much better
  description of the surface brightness of elliptical galaxies over a wide
  range in luminosity \citep{Graham05}. Here $\Sigma_{\mathrm{50}} \equiv
  \Sigma(\SersicR)$ is the amplitude of the profile at $\SersicR$, the radius
  enclosing half the mass (or light) in projection, and $n$ is the \Sersic{}
  index, which sets the concentration of the profile ($n=4$ corresponds to an
  $R^{1/4}$ profile and $n=1$ to an exponential profile).  C13 found that
  $\SersicR$ and $n$ for the central galaxies of massive haloes in their
  simulation matched observations well -- both parameters were found to
  increase systematically with halo mass.  For massive elliptical galaxies, the
  increase in size and \Sersic{} index with mass is driven by accretion, as in
  the models of \cite{Cole:2000aa} and \citet{Naab09b}.  The Phoenix BCGs represent
  extreme examples of these trends -- not only are they more massive and
  dynamically younger than other elliptical galaxies, but almost all of their
  stars are accreted. This results in large sizes and high \Sersic{} indices. 

Single \Sersic{} functions are a reasonable first order description of the
BCG+ICL stellar mass surface density profile in all nine Phoenix clusters, but
they are far from perfect.  The parameters of maximum likelihood \Sersic{} fits
are given in Table~\ref{tab:phoenixprofiles}. For these fits, we excluded the
inner 5~kpc, because numerical softening dominates at $\lesssim 1$~kpc and
because we neglect the gravity of baryons, which is likely to be a poor
approximation in that region\footnote{Observers also exclude the central
regions when fitting surface brightness: \citet{Seigar07} excluded the
innermost 2--5~kpc to avoid regions dominated by PSF deconvolution, and
\citet{Zibetti05} excluded the inner 10--20~kpc. Since these profiles are
concentrated, the size of the excluded region can influence the ratio of
components in a double \Sersic{} fit.}.

A number of authors, in particular \cite{Seigar07}, advocate fitting observed
BCG profiles with the sum of two \Sersic{} components; the outer component was
found to have an exponential form ($n\sim1$) in many cases (see also
\citealt{Donzelli11}).  Fig.~\ref{fig:sersicprof} and columns 8--13 of
Table~\ref{tab:phoenixprofiles} show the results of double \Sersic{} fits to
our simulations, with the constraints given above.  The residuals for these
fits are $\lesssim 50\%$ over three orders of magnitude in radius. In five
cases (Ph-B, E, G, H and I) we find that the \Sersic{} component with larger
half-mass radius (all $\gtrsim 250$~kpc) has $n\lesssim 2$ (for Ph-G and Ph-H,
$n<1$). \CD{The} haloes with exponential outer components are also,
\CD{perhaps} not surprisingly, those with the strongest cD-like departures
from $R^{1/4}$ profiles. \CE{These simulation results provide a possible
explanation for the exponential outer components found by} \cite{Seigar07} and
\citet{Donzelli11}.

C13 concluded that double \Sersic{} profiles are also an excellent fit to the
average profile of central elliptical galaxies in haloes $M_{200}\lesssim
10^{14} \, \Msol$. However, there is an important difference between that
statement and our conclusions regarding the most massive BCGs. C13 showed that
the double \Sersic{} form in their central galaxies was due to the gradient in
the ratio between in situ stars (dominant at $R \lesssim 100$~kpc) and accreted
stars. In contrast, as we show in the next section, the double \Sersic{} form
of the Phoenix BCGs is driven by a transition between \textit{different
accreted components}, in different states of dynamical relaxation and/or
symmetry around the BCG. The transition between in situ and accreted stars is
much less conspicuous.

\section{Surface density substructure}
\label{sec:substructure}

Previous figures have shown that the azimuthally averaged surface density
profiles of accreted stars in the BCGs of massive clusters have similar shape
and amplitude. We now investigate why this is the case. We start from the fact
that a BCG density profile can be considered as the superposition of many
profiles, each corresponding to stars accreted from a single
progenitor\footnote{ For this section, the progenitor of a given star particle
is the DM halo (or subhalo) to which it was last bound before joining
the main branch of the cluster merger tree. Each of these progenitor haloes is
the root of its own merger tree and hence will contain stars formed in many
different galaxies. Note that the mass contributed \textit{by} a progenitor to
the BCG need not equal the mass \textit{of} the progenitor; many of the most
significant progenitors survive and retain a significant fraction of their
stellar mass at $z=0$.} galaxy.

\subsection{Surface brightness of debris components}
\label{sec:contrib}

\begin{figure} 
  

  \includegraphics[width=84mm, trim=0.0cm 0cm 0.1cm 0cm, clip=True]{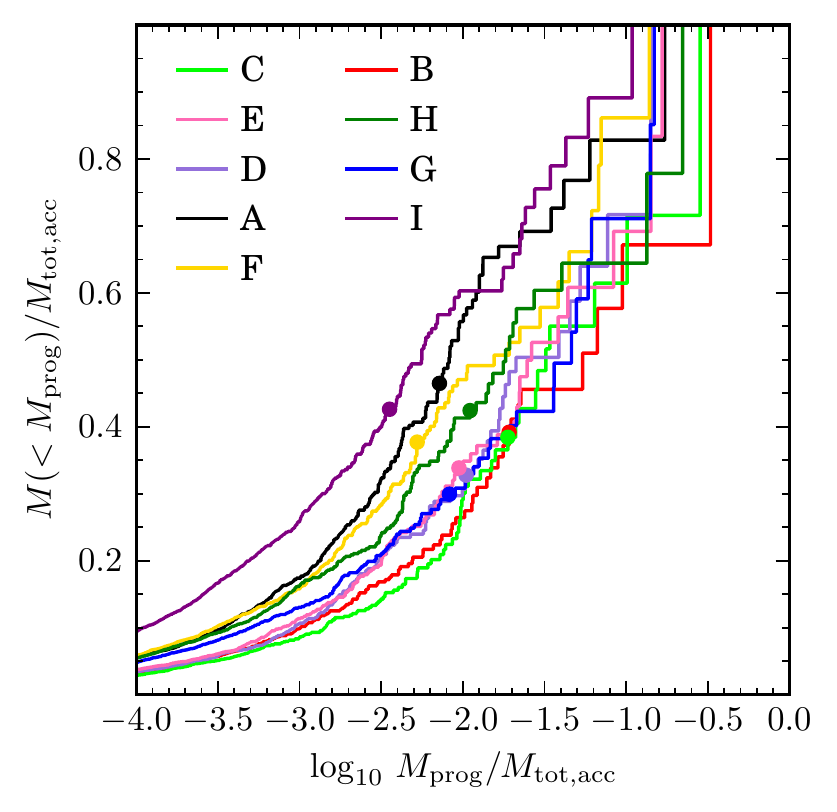}

  \caption{\CD{Distribution of progenitor stellar mass contributions} for
	  each Phoenix BCG, showing the fraction of the total accreted stellar
	  mass, $M_{\mathrm{tot, acc}}$ (vertical axis), from systems with
	  stellar mass, $M_{\mathrm{prog}}$ less than a given fraction of
	  $M_{\mathrm{tot, acc}}$ (horizontal axis). The large steps at high
	  $M_{\mathrm{prog}}$ correspond to the most massive contributions from
	  individual progenitors. Dots mark the mass fraction corresponding to
	  $\Nprog$, an indication of the number of significant progenitors (see
	  text and Table~\ref{tab:phoenixstars}). The Phoenix BCGs have many
	  more significant progenitors than the less massive galaxies studied
	  in C13 and \citet{Cooper10}.} 

\label{fig:contrib_spectrum}

\end{figure}

\begin{figure}
\includegraphics[width=84mm, trim=0.0cm 0cm 0.1cm 0cm, clip=True]{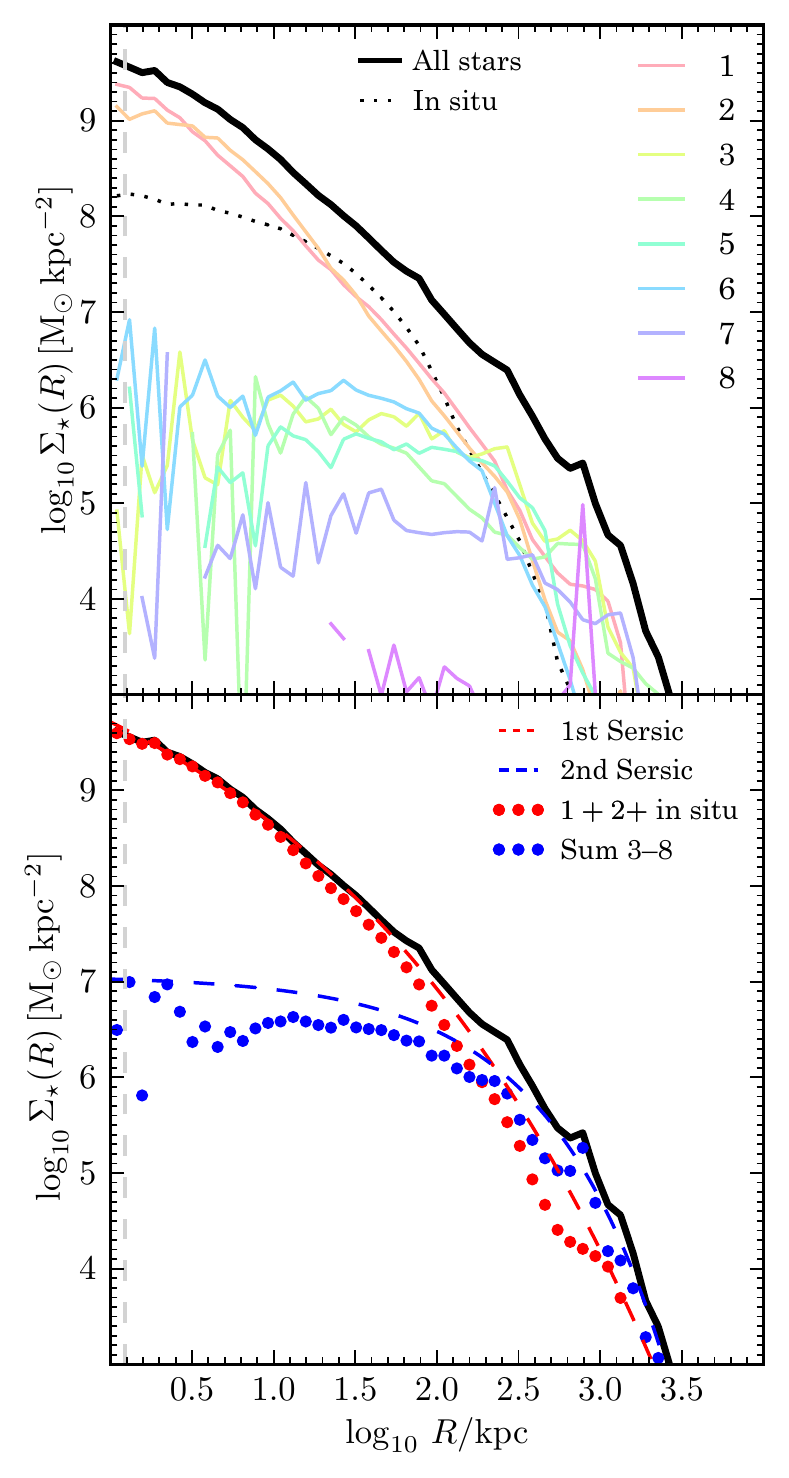}

\caption{Decomposition of the stellar mass surface density profile of Ph-E
	(thick black line) into components corresponding to the eight accreted
	systems that contribute most mass to the BCG/ICL (thin coloured lines,
	labelled in rank order of stellar mass contributed, from 1 to 8). The
	in situ component is shown by the black dotted line. In the bottom
	panel, red and blue dashed lines show the components of a double
	\Sersic{} fit to the total stellar density profile. A vertical grey
	dashed line marks the softening length.} 

\label{fig:contribs} 
\end{figure}

Fig.~\ref{fig:contrib_spectrum} shows the stellar mass of stripped debris that
each individual progenitor contributes to the BCG, \CD{in rank order}. The
largest (rightmost) step in each of the curves in
Fig.~\ref{fig:contrib_spectrum} corresponds to the largest single contribution,
which accounts for only $10$--$35$ per cent of the total accreted stellar mass
($M_{\mathrm{tot, acc}}$). Most progenitor galaxies contribute much smaller
fractions. $\Nn$, the number of progenitors taken in decreasing mass order
required to account for 90 per cent of $M_{\mathrm{tot, acc}}$, spans the range
$61 \leq \Nn \leq 657$ (Table~\ref{tab:phoenixstars}). \CD{\citet{Cooper10}
took} the \CA{square root of the} second moment of the distribution in
Fig.~\ref{fig:contrib_spectrum} (labelled $\Nprog$) as representative of the
number of `significant' progenitors.  For our clusters we find $8 \lesssim
\Nprog \lesssim 43$.  Depending on the halo, $50$--$70$ per cent of
$M_{\mathrm{tot, acc}}$ is accounted for by these significant progenitors, with
individual masses $M_{\mathrm{prog}} \gtrsim 0.3$--$3$ per cent of the total
(hence roughly $\gtrsim10^{11} \Msol$ each). $\Nprog$ is therefore larger for
massive clusters than for the stellar haloes \CC{of} Milky Way-like
systems, where the bulk of accreted mass is contributed by fewer than five
progenitors, typically with only one or two dominating \citep{Cooper10}. 

The top panel of Fig.~\ref{fig:contribs} shows how the overall accreted density
profile is built up by these contributions, using Ph-E as an example (the
bottom panel is discussed in section~\ref{sec:twocomponents}). Separate
profiles are shown for debris from each of the top eight most massive progenitor
galaxies (which make up $\sim60$ per cent of the accreted mass according to
Fig.~\ref{fig:contrib_spectrum}). Two types of profiles can be distinguished in
this figure. The profiles of stars from the two most massive progenitors
(labelled 1 and 2) are centrally concentrated, with similar shapes that
resemble the overall profile. Debris clouds from the other progenitors have
much larger effective radii, lower concentration (\Sersic{} index) and hence
very little mass within $R\lesssim 100 \, \mathrm{kpc}$. In the case of
progenitor 8, the debris is concentrated almost entirely in one radial
bin at $\sim800$~kpc.

\begin{figure*}
\includegraphics[width=0.9\textwidth, trim=0.25cm 0.25cm 0.25cm 0.25cm, clip=True]{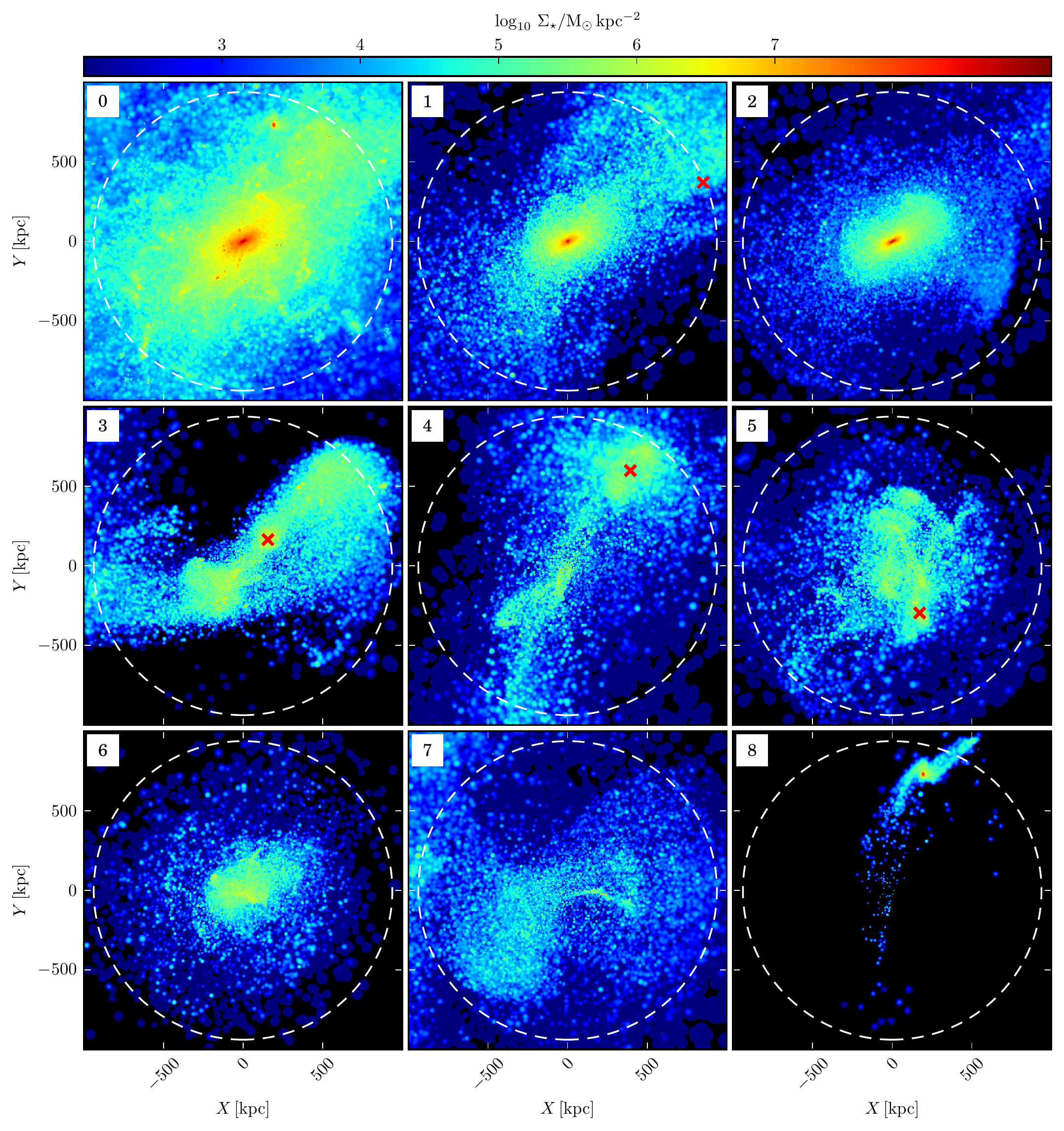}

\caption{Surface mass density images as in Fig.~\ref{fig:gallery} for stars
associated with each of the top eight accreted components in Ph-E (labelled from 1
to 8 in in order of mass contributed). The white dashed circle shows
$\frac{1}{2}R_{200}$. \CE{Red crosses mark the centre of the parent subhalo of
each progenitor in cases where it survives}. Panel 0 shows the total surface
density of all stars \CX{bound to the main halo (i.e. excluding stars bound
to satellites)}.  The most significant debris components cover a wide range in
terms of central concentration, symmetry and dynamical relaxation.} 

\label{fig:contribs_gallery} \end{figure*}

Fig.~\ref{fig:contribs_gallery} shows images of each component, which help to
explain the differences between their profiles.  Components 1 and 2 are
centrally concentrated and have a smooth distribution in projection, with only
faint asymmetric structure to suggest they were accreted. These two galaxies
fell into the cluster at $z\approx1.0$ and $z\approx1.8$ respectively.
\CD{Component 1 was disrupted very rapidly, although a very small remnant
core survives as a subhalo at $z=0$ (marked by a red cross). Stars were
stripped more gradually from component 2 over a period of 4~Gyr, but no core
survives at $z=0$}. Together these progenitors contribute $\sim30$ per cent of
the accreted stellar mass. 

The progenitors of components 3, 4 and 5 fell into the cluster at redshifts
$0.7 < z < 1.4$, but were stripped more recently \CA{($z\lesssim0.5$)}, with
only half of the stars they contribute to the BCG being stripped earlier than,
respectively, 2, 2 and 5~Gyr ago. The progenitor of component 3 spent
nearly 6~Gyr in the cluster potential before losing any stars to tidal
stripping. All these components have an unrelaxed morphology, with debris
tracing stream-like orbits. Component 3 shows the early stages of radial shell
formation \citep[e.g.][]{Cooper11b} and component 5 traces a rosette-like
orbit, seen edge-on in this projection. Each of these components is associated
with a surviving resolved subhalo at $z=0$. \CE{Unlike the inconspicuous
remnant core of progenitor 1, these subhaloes correspond to some of the most
luminous cluster members in the top left panel of
Fig.~\ref{fig:contribs_gallery}.}

Progenitor 6 fell in and was disrupted over a similar time-scale to progenitor
2. Although its debris appears quite uniformly distributed it is much less
centrally concentrated. The diffuse shell-like debris of progenitor 7 is the
result of very rapid disruption (over $\lesssim1$~Gyr) after infall at
$z\sim0.7$.  Neither progenitor survives at $z=0$. Finally, progenitor 8
appears very much like an intact galaxy with tidal tails. This is partly a
projection effect; most of these stars are turning around in a `kink' at the
apocentre of a complex, approximately figure-of-eight orbit, the plane of which
is almost perpendicular to the line of sight. Despite \CX{the apparent
concentration of stars in this projection, the DM particles from this
progenitor are extremely diffuse and so are assigned to the main halo by
\subfind, rather than to a self-bound subhalo (note that the body of this
debris cloud is $\sim100$~kpc in extent, and that, by chance, its central
concentration -- the red region on the image -- is dominated by just two
massive stellar tags)}. This relatively massive galaxy is identified as a
`survivor' with a sub-resolution halo in our semi-analytic model, making this
one of the rare cases where the \CX{unbound} particles from such a galaxy are
still confined to a small region in configuration space.  

We conclude that the different profiles of the debris components reflect a
range of \CB{possibilities} for how progenitor galaxies can be disrupted,
varying according to their mass, accretion time and orbit. We can infer that
centrally concentrated profiles with high \Sersic{} index correspond to systems
that sink (rapidly) to the centre of the cluster and/or that are accreted at $z
\gtrsim 1$. These tend to include at least some of the most massive
progenitors: components labelled 1, 2 and 6 in the above figures are `major'
mergers, having stellar mass ratios with the BCG of $\sim 1:3$ at infall.  Such
mergers are likely to result in violent relaxation \CX{that erases phase space
structure in high density regions, although a small but significant fraction of
stars from the same events can also be deposited at much larger radii and
remain in coherent structures such as shells. In cases where the progenitor
satellite has a lower mass ratio (typically 1:10 or higher) and is heavily
stripped on a weakly bound, low-eccentricity orbit, the majority of the debris
is deposited at cluster-centric radii $\gtrsim100$~kpc, forming an extended,
diffuse profile. Such accretion events are increasingly common at low redshift,
and longer dynamical times naturally preserve stream and shell structure in
recently accreted debris at these radii}. In general, the massive debris
components associated with diffuse profiles are also visually `unrelaxed',
although the degree of visual substructure is not always obvious from the
concentration or smoothness of the profile.

We find different relative proportions of `relaxed' and `unrelaxed' components
among the most significant progenitors across our nine simulations.  In general,
however, progenitors with diffuse profiles dominate the outskirts and smooth
concentrated profiles dominate in the centre of the cluster, as in the example
shown.  Individual counterexamples are not hard to find -- dynamically young
structures can be found in the centre of some clusters, and there is an
appreciable nett contribution at large radii from components with a `smooth'
spatial distribution.

\subsection{Physical origin of double \Sersic{} profiles}
\label{sec:twocomponents}

We now return to the meaning of the two components picked out by a double
\Sersic{} fit to the total surface density in
Section~\ref{sec:functional_forms}. These are shown for Ph-E by red and blue
dashed lines in the bottom panel of Fig.~\ref{fig:contribs}. Their mass ratio
is approximately 2:1, with the more concentrated component (red line) being the
more massive. The transition between the two at $\sim 200$~kpc corresponds
roughly to the departure from an $R^{1/4}$ law in Fig.~\ref{fig:devprof} (the
inner component has $n\approx5$). We see a clear similarity between these two
components and the sum of the profiles of the `relaxed' and `unrelaxed'
progenitor profiles respectively.  We conclude that the large-scale inflection
in the shape of the overall profile results from a transition between inner
regions dominated by `relaxed' accreted components and outer regions dominated
by `unrelaxed' accreted components. In Ph-E, this transition is picked out,
approximately, by the double \Sersic{} fit. 

This result holds across our sample of nine haloes, even though decompositions of
the density profile show a different mix of relaxed and unrelaxed
sub-components in each case.  Where a double \Sersic{} profile is not strongly
favoured over a single \Sersic{} (e.g.  Ph-C, Ph-F), plots equivalent to
Fig.~\ref{fig:contribs} show a continuum of profiles, varying from `relaxed' to
`unrelaxed' with increasing radius. Where a double \Sersic{} fit is strongly
favoured, the component with larger effective radius and lower \Sersic{} index
provides a reasonable estimate of the cumulative contribution of `unrelaxed'
accreted debris. \CX{A similar contrast in dynamical state has been found in
hydrodynamical cluster simulations when stars are separated into discrete
components according to their binding energy and kinematics \citep[see e.g.
fig.~7 of][]{Cui14a}.}

Substantial mass in one or more `unrelaxed' debris components appears to be the
origin of `cD envelopes' in our simulated clusters. In particular, the diffuse
light around other very massive galaxies in the cluster is responsible for the
majority of the strong `cD envelope' effect seen in haloes Ph-B, Ph-G and Ph-H.
The distorted outer envelopes of these bright cluster members (see
Fig.~\ref{fig:gallery}) are not concentric with the BCG, hence their
contributions to the overall profile are less concentrated (i.e. have lower
\Sersic{} index) when measured in BCG-centred apertures. In Ph-E and PH-I, on
the other hand, the envelope is not dominated by a single progenitor and is
concentric with the BCG. \CE{The choice of the BCG can be ambiguous during
cluster mergers}. \CD{When} several BCG candidates lie near the centre of a
unrelaxed cluster, as in Ph-B and Ph-G, their envelopes may be much larger than
their separation and hence the projected centroid of the diffuse light
\CD{might} not correspond \CD{to any of the brightest} galaxies.

\section{ICL fraction}
\label{sec:iclfraction}

\begin{figure}
  \includegraphics[width=84mm, trim=0.0cm 0cm 0.1cm 0cm, clip=True]{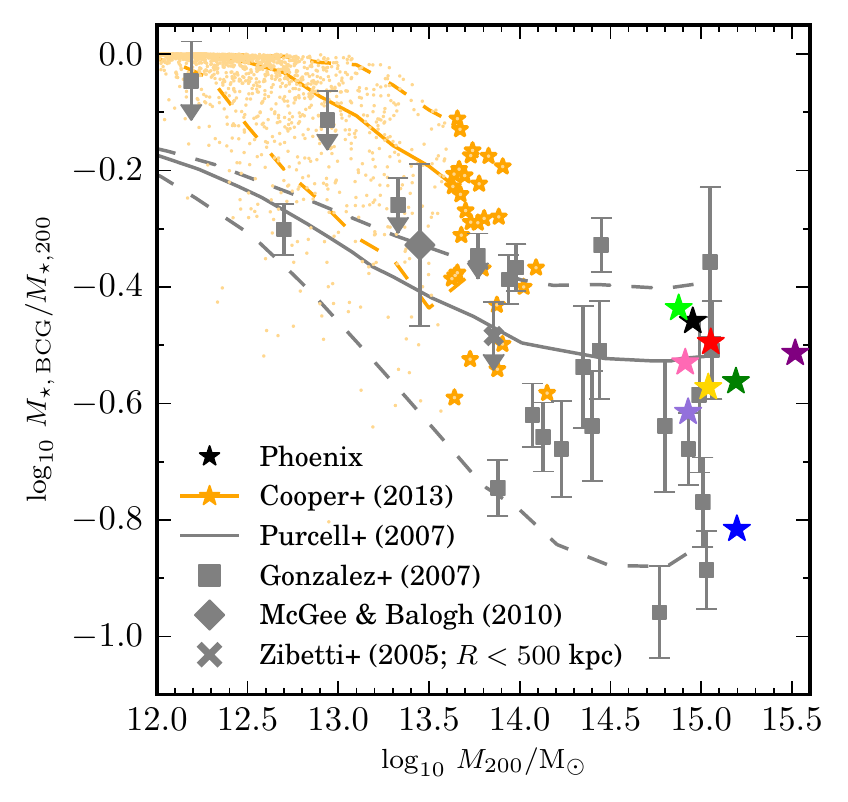}

  \caption{Stellar mass of central galaxies in our models \CB{(with no
      separation of ICL)} within $R_{200}$, as a fraction of the total stellar mass within
	  $R_{200}$ \CD{(this total mass includes all satellite galaxies)}.
	  Filled star symbols correspond to the individual Phoenix haloes with
	  our fiducial model that includes stars from sub-resolution
	  semi-analytic haloes in the BCG+ICL.  Orange points show galaxies in Millennium
	  II from C13; individual haloes of $M_{200} >
	  10^{13.5}\mathrm{M_{\sun}}$ are highlighted with orange star
	  symbols. Orange lines show the median (solid) and 10--90 per cent
      range (dashed) of the Millennium II data. \CC{Grey} points show
	  observational data from \citet[][squares, assuming
	  $M_{\star}/L=3.6$]{Gonzalez07}, \citet[][cross]{Zibetti05} and
	  \citet[][diamond]{McGee10}. Arrows indicate measurements in an
	  aperture of $R<R_{200}$, hence upper limits. Grey lines show the
	  median and 10--90 per cent range for the model of \citet{Purcell07}.}

  \label{fig:icl_fraction}
\end{figure}

\begin{figure}
  \includegraphics[width=84mm, trim=0.0cm 0cm 0.0cm 0cm, clip=True]{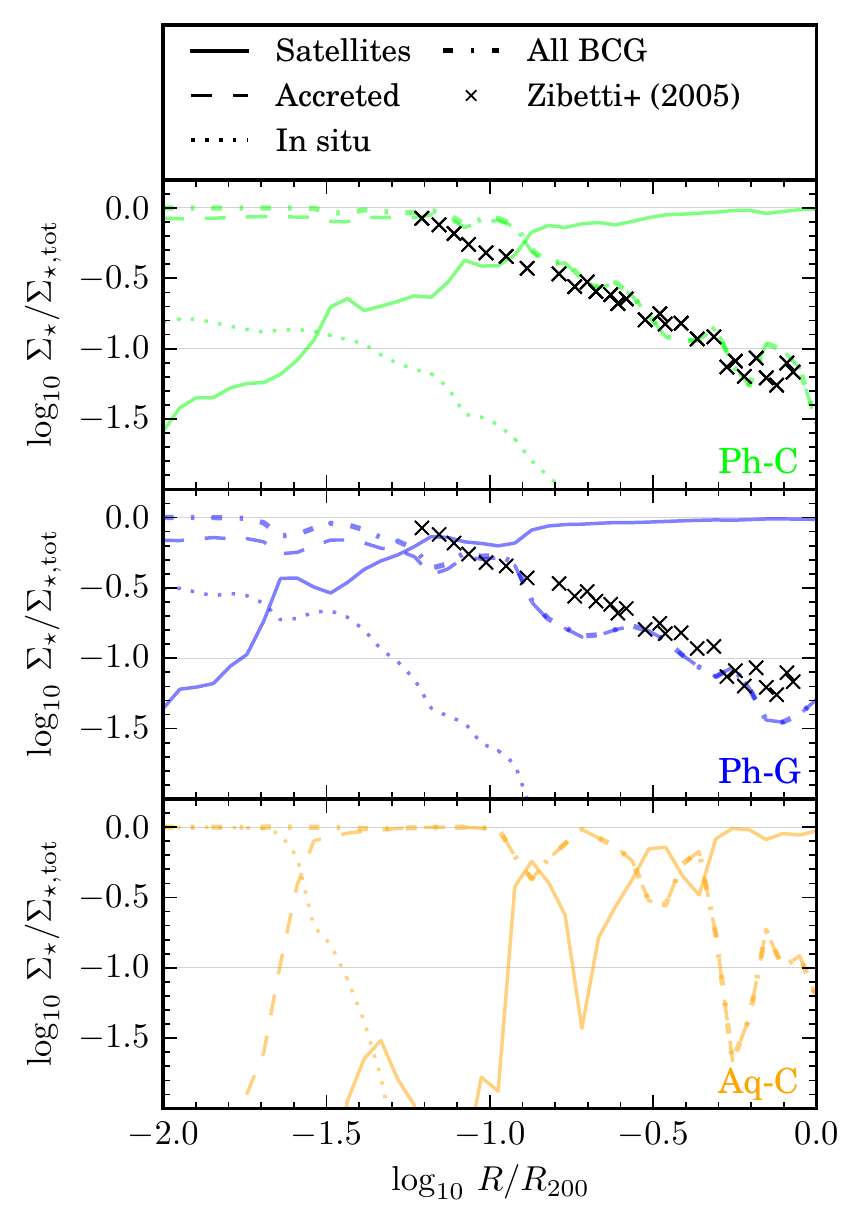}

  \caption{Surface mass density fraction of stars in the BCG (subdivided into
  accreted and in situ) and satellites, for haloes Ph-C (top) and Ph-G
  (middle). Crosses show the measurement of ICL mass fraction by
  \citet{Zibetti05}. Satellite stars account for most of the stellar mass
  beyond $0.1 \, R_{200}$ and a large fraction at smaller radii. The lower
  panel compares to one of the Aquarius Milky Way-mass haloes (Aq-C), in which
  the central galaxy and its stellar halo dominate out to $0.5 \, R_{200}$.}

  \label{fig:icl_ratio_radius}
\end{figure}

\begin{figure}
  \includegraphics[width=84mm, trim=0.0cm 0cm 0.0cm 0cm, clip=True]{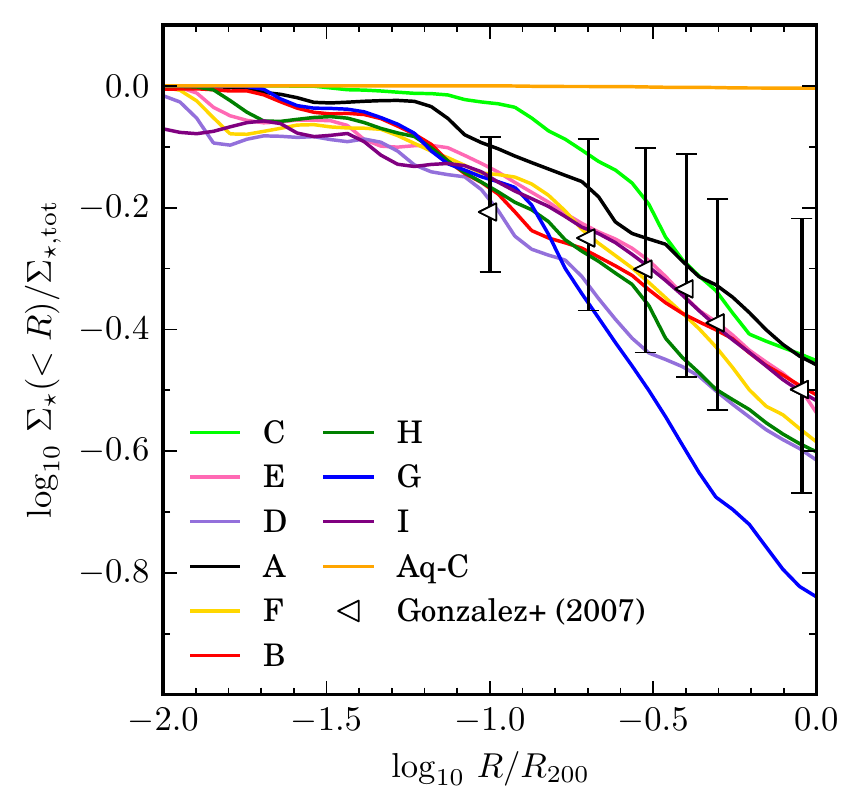}

  \caption{\CB{The cumulative enclosed mass fraction in BCG+ICL
  stars at different radii. Triangles show data from \citet{Gonzalez07};} solid
  lines show the same in Phoenix. The orange line corresponds to Aquarius Milky
  Way-mass Aq-C, in which the central galaxy and its halo dominate the total
  mass. }

  \label{fig:icl_ratio_radius_cumulative}
\end{figure}

The fraction of the stellar mass of a cluster made up by intracluster stars has
been studied by many authors \citep[e.g.][]{Thuan77, Bernstein95}. This
measurement depends sensitively on the nature of the BCG and ICL density
distribution. Recent observational estimates of the `ICL fraction' range from
10 to 50 per cent \citep{Gonzalez07, McGee10, Sand11}. This wide range may be
the result of lumping together very different sample selections (cluster and
galaxy masses) and definitions of ICL, alongside scatter caused by
observational uncertainties and stochastic variations between clusters
\citep[as discussed by][]{Lin04}. We compare our simulations with
\citet{Zibetti05} and \citet{Gonzalez07}, because both these studies are based
on large, well-defined samples of galaxies and do not depend strongly on
assumptions about the nature or distribution of the ICL component. 

Fig.~\ref{fig:icl_fraction} plots data from \citet[][]{Gonzalez07}, who
measured the ratio between the combined mass of the BCG and ICL (which we call
$M_{\star \mathrm{BCG}}$) and the total stellar mass within $R_{200}$,
$M_{\star, \mathrm{200}}$ (this includes stars in satellite
galaxies)\footnote{We are grateful to A. Gonzalez for providing measurements in
an aperture of $R_{200}$ rather than $R_{500}$ as given in \citet{Gonzalez07};
halo masses have been scaled assuming an NFW concentration of 5, hence by a
factor of $M_{200}/M_{500} = 1.38$.}.  Our Phoenix cluster simulations cover a
range $15 < M_{\star \mathrm{BCG}}/M_{\star, \mathrm{tot}}< 40$ per cent and
therefore agree well with the spread of the \citeauthor{Gonzalez07} data at
comparable $M_{200}$.  \CB{If stars associated with sub-resolution haloes are
treated as bound to satellites rather than the BCG, the Phoenix BCG stellar
mass fractions are reduced by $\lesssim0.2$~dex}.

\citeauthor{Gonzalez07} find a trend with $M_{200}$, such that the BCG and its
stellar halo account for $\sim50$ per cent of the total stellar mass at
$M_{200}\sim10^{13.5}\Msol$ (although see \citealt{Balogh08}).  \citet{McGee10}
obtained a similar value based on observations of intergalactic supernovae in
galaxy groups ($M_{200}\sim10^{13.5}$, black circle; see also
\citealt[][]{GalYam03, Sand11}). \citet{Zibetti05} found a mass fraction of 33
per cent in haloes of average mass $\sim7\times10^{13} \Msol$. These
measurements are for haloes much less massive than those of Phoenix, so
Fig.~\ref{fig:icl_fraction} compares them with the Millennium II results of C13
(orange stars and lines). At $M_{200}=10^{13.5}\Msol$ we find a median value of
$M_{\star, \mathrm{BCG}}/M_{\star,200}=0.6$, broadly consistent with the data but
implying a more gradual decline with increasing $M_{200}$. Improving the
statistics of observations in this mass range (for example through stacking;
\CE{C13; \citealt{Budzynski14, DSouza14_arxiv}}) would be beneficial,
as would further simulations of low-mass galaxy clusters ($10^{13} < M_{200}
<10^{15}\Msol$).

\citet{Purcell07} used simple scaling relations to populate haloes from an
$N$-body simulation with stars and followed their merging histories to infer
`BCG+ICL' stellar mass fractions \citep[see also][]{Lin04,Conroy07}. Their
predictions for the distribution of $M_{\star \mathrm{BCG}}/M_{\star,
\mathrm{tot}}$ and its variation with $M_{200}$ are shown by the grey lines in
Fig.~\ref{fig:icl_fraction}. The \citeauthor{Purcell07} model agrees with our
direct simulations at $M_{200} \gtrsim 10^{14.5} \Msol$ but predicts a
substantially smaller fraction of stellar mass in the BCG at $M_{200} \lesssim
10^{13.5} \Msol$. Although this is compatible with one data point of
\citet{Gonzalez07}, Abell 3166, it is inconsistent with observations of the
Milky Way and M31 (and similar galaxies), which have relatively
well constrained halo masses and BCG+ICL fractions of $\gtrsim 90\%$
\citep{Helmi08, Li08_mwmass, McConnachie:2012aa, MD10}. In this respect the
Millennium II results of C13 are more consistent with the data.

\subsection{Radial variation}

Fig.~\ref{fig:icl_ratio_radius} shows how $M_{\star
\mathrm{BCG}}/M_{\mathrm{200}}$ varies with radius for haloes Ph-C and Ph-G
(dot--dashed lines). Radii are expressed as a fraction of $R_{200}$.  We break
each of these curves into separate components representing in situ BCG stars
(dotted lines), accreted BCG stars (dashed lines) and stars in satellite
galaxies (solid lines).

In Ph-C, BCG stars account for more than 90 per cent of the stellar mass
projected in annuli $R \lesssim 0.1 \: R_{200}$; their contribution falls to
$\sim3$ per cent at $R_{200}$. These findings hold on average for the nine
clusters, with scatter comparable to the fluctuations in individual profiles.
As a result, all the simulations (except Ph-G, see below) are in very good
agreement with the average behaviour of the \citet{Zibetti05} stack of BCGs
(crosses). This agreement is remarkably insensitive to the $\sim1.5$~dex
variation in $M_{200}$ across our sample. 

For comparison, the lower panel of Fig.~\ref{fig:icl_ratio_radius} shows a
high-resolution simulation of a Milky Way-like halo analogue with
$M_{200}\sim10^{12} \, \mathrm{M_{\sun}}$ (Aquarius-C-2, \citealt{Cooper10}).
The trends in the individual components are similar to those in BCGs, with
satellite stars dominating the projected mass at radii beyond 10 per cent of
$R_{200}$.  However, the Aquarius profile shows much larger
\CB{oscillations} due to individual stellar streams and satellites. Even
though satellite stars dominate locally over $\sim50$ per cent of the projected
area within $R_{200}$, they account for $\lesssim$10 per cent of the total
stellar mass, as shown in Fig.~\ref{fig:icl_fraction}.

Finally, in Fig.~\ref{fig:icl_ratio_radius_cumulative}, we examine the
\textit{cumulative} value of  $M_{\star \mathrm{BCG}}/M_{\star, \mathrm{tot}}$
within a given fraction of $R_{200}$. We \CB{find good agreement with the
data of \citet{Gonzalez07}, who also found that the value of this ratio at any
given fraction of $R_{200}$ scales with cluster velocity dispersion (their
fig.~5).  Fig.~\ref{fig:icl_ratio_radius_cumulative} supports such a trend in
our models, albeit weakly, despite our $M_{200}$ range being narrower than that
of \citet{Gonzalez07}.}

Ph-G deviates most from the average behaviour of $M_{\star,
\mathrm{BCG}}/M_{\star, \mathrm{tot}}$ in
Figs.~\ref{fig:icl_fraction}--\ref{fig:icl_ratio_radius_cumulative}. The latter
two figures show that the ratio of BCG stars to satellite stars is lower at all
radii compared to the observational average of \citet{Zibetti05}.  Several of
our haloes have comparable or higher satellite mass fractions in their inner
regions, hence it is the low BCG mass ratio at $R \gtrsim 0.1R_{200}$ that
makes Ph-G an outlier in Fig.~\ref{fig:icl_fraction}.  Inspection of
Table~\ref{tab:phoenixstars} shows that the ratio of BCG stellar mass to total
halo mass, $M_{\star,\mathrm{BCG}}/M_{200}$, is particularly low in Ph-G, so it
appears to be `missing' stellar mass from its outer regions. 

The most obvious reason for this deficiency is that Ph-G is dynamically very
young. It is essentially two clusters in one halo (see Fig.~\ref{fig:gallery}),
where the massive satellites have yet to suffer the full effects of the newly
combined DM potential and the two BCGs have yet to merge \citep[see
also][]{Dolag10}. As yet, however, there is little direct evidence for recent
mergers driving the scatter in the observations plotted in
Fig.~\ref{fig:icl_fraction}. Of the three most comparable data points of
\citet{Gonzalez07} \CB{in Fig.~\ref{fig:icl_fraction}}, Abell clusters 2721
and 3705 have no dominant BCG (i.e.  Bautz-Morgan type III) and Abell 3750 also
has a strongly bimodal X-ray morphology \citep{Sivanandam09}, both of which are
suggestive of unrelaxed clusters. Abell 1651, however, has symmetrical X-ray
contours and Bautz-Morgan type I-II \citep{Sivanandam09}.

\subsection{BCG contribution to surface density}

\begin{figure}
  \includegraphics[width=84mm, trim=0.0cm 0.2cm 0.0cm 0.2cm, clip=True]{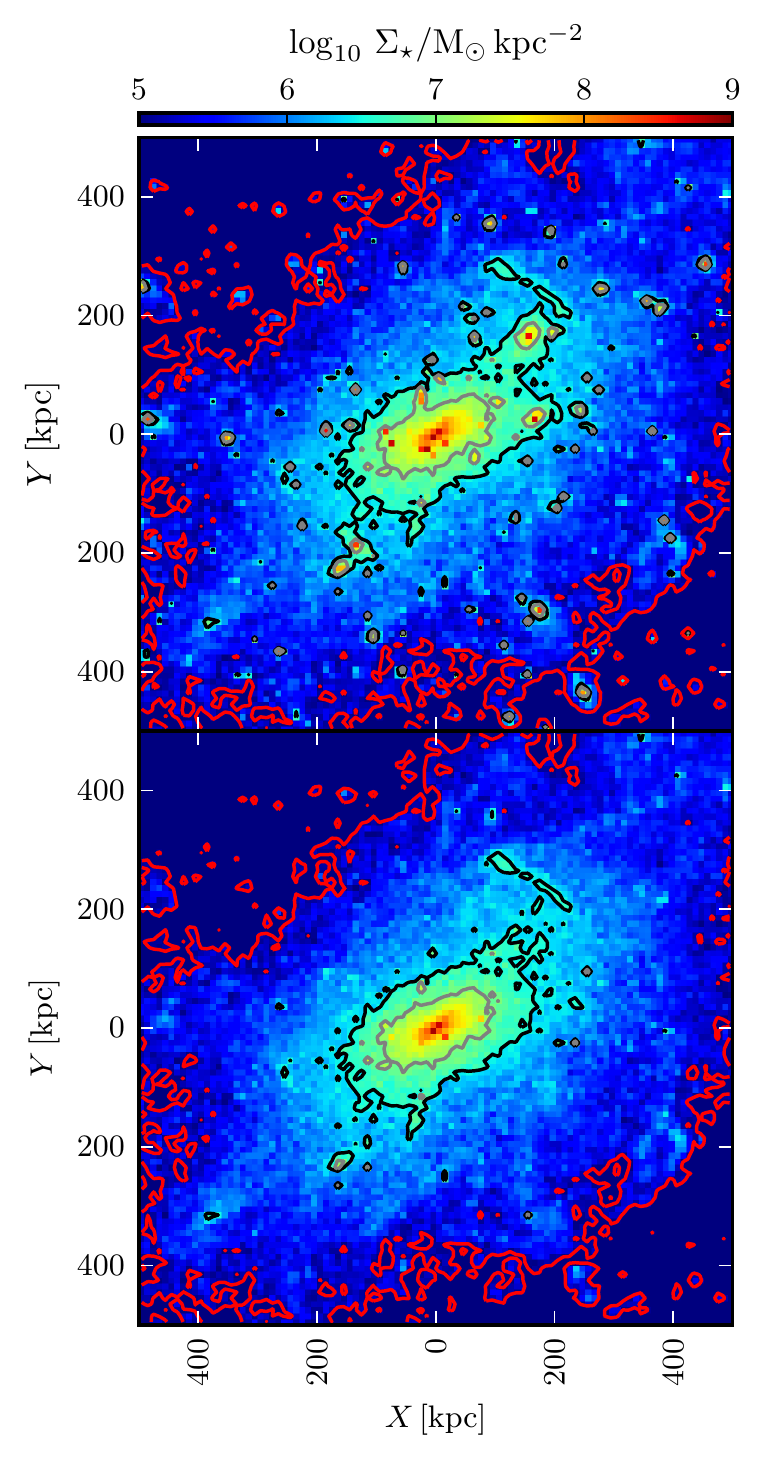}

  \caption{Central region of Ph-E at a resolution of 1~$\mathrm{kpc^{2}}$ per
  pixel. Contours mark stellar mass surface densities $10^{5}$ (red),
  $10^{6.5}$ (black) and $10^{7}$ (grey) $\sdunits$. The top panel shows all
  stellar mass in the simulation, including cluster members other than the BCG.
  In the bottom panel, only BCG stars are shown. BCG stars drive the
  orientation, extent and amplitude of the diffuse light at $\Sigma \lesssim
  10^{6.5} \sdunits$.}

  \label{fig:pixel_stats_pic}
\end{figure}

\begin{figure}
  \includegraphics[width=84mm, trim=0.0cm 0.2cm 0.0cm 0.25cm, clip=True]{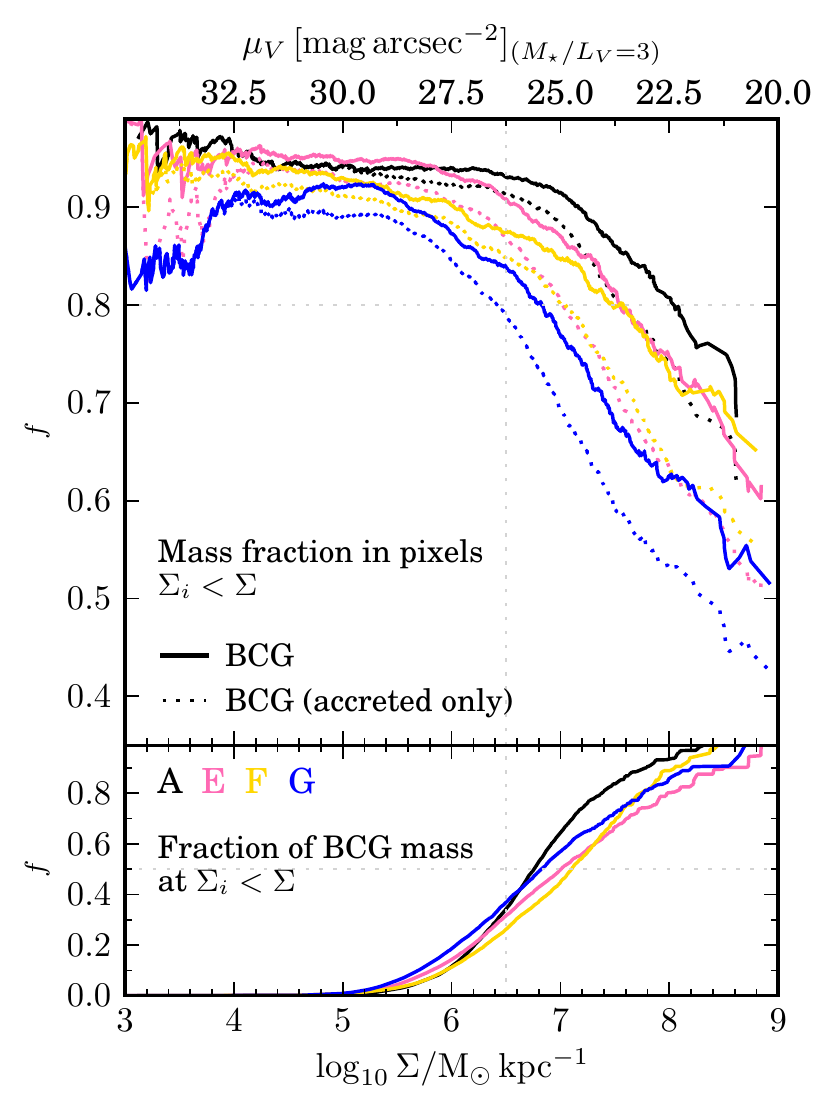}

  \caption{Top panel: \CX{Fraction of the total stellar mass in low surface
  mass density pixels which is assigned to the BGC as a function of $\Sigma$,
  the upper surface mass density limit.} Upper axis converts $\Sigma$ to
  $V$-band surface brightness assuming a mass-to-light ratio of 3. Dotted lines
  exclude in situ \CX{BCG} stars. Bottom panel: fraction of total BCG stellar
  mass in 1~$\mathrm{kpc^{2}}$ pixels with surface density less than $\Sigma$.
  The $\Sigma = 10^{6.5} \sdunits$ contour is a good empirical threshold for
  the diffuse BCG envelope: selecting all pixels below this density maximises
  the fraction of light per pixel due to the BCG ($\gtrsim80$ per cent) while
  also recovering $\sim30$ per cent of its total stellar mass.}

   \label{fig:pixel_stats}
\end{figure}

A surface brightness threshold is sometimes applied to images to separate ICL
from the light of BCGs and/or other cluster galaxies (typically in the range
$25$--$27 \sbunits$; \citealt{Zibetti05,Rudick11}). Our models can be used to
determine how `efficient' such cuts are, in terms of the fraction of the total
accreted BCG mass they recover, and how `pure' they are, in terms of the
fraction of recovered light per pixel that is really due to the BCG rather than
faint cluster galaxies.

The top panel of Fig.~\ref{fig:pixel_stats_pic} shows a simple `SDSS like' mock
observation: a $1\,\mathrm{Mpc^2}$ image of the centre of cluster Ph-E with
$1\,\mathrm{kpc^2}$ pixels.  The lower panel shows only those stars associated
with the BCG as we define it. Fig.~\ref{fig:pixel_stats} shows the cumulative
fraction of stellar mass in all pixels below a given surface brightness that is
associated with the BCG. The remainder is associated with other cluster
galaxies. The bottom panel shows the fraction of the BCG's total mass in pixels
below a given surface density. 

For an image such as Fig.~\ref{fig:pixel_stats_pic}, $90$ to $95$ per cent of
the stellar mass in pixels with $\Sigma < 10^{5} \sdunits$ belongs to the BCG,
but this accounts for $\lesssim 1$~per cent of its total stellar mass. Over
$80$ per cent of the mass in pixels with a stellar mass surface density
$\Sigma < 10^{6.5} \sdunits$ is associated with the BCG.  Almost all of this is
accreted (compare dotted and solid lines).  The in situ contribution to the BCG
increases at higher $\Sigma$, as does the contribution from other cluster
galaxies. 

A threshold of $\Sigma \lesssim 10^{6.5} \sdunits$ ($\mu_{V} \gtrsim 26.5$) is
a reasonable compromise, selecting $\sim30$ per cent of the BCG mass at
$\gtrsim 80$ per cent purity. \CE{Hence our simulations predict that
approximately $20$~per cent of the luminosity in pixels fainter than this
threshold is associated with surviving galaxies. This prediction can be compared
to measurements of the luminosity contributed by unresolved galaxies and the
low surface brightness regions of resolved galaxies in real images}. If stars
associated with galaxies with sub-resolution haloes are \CX{not} counted towards
the BCG mass, the purity for $\Sigma \lesssim 10^{6.5} \sdunits$ becomes 50 to
70 per cent (see Appendix A).  \CD{This suggests that the distribution of pixel
surface brightness in deep photometric observations might be a useful way to
constrain theoretical models of galaxy disruption and stellar stripping.}

\section{Discussion: a separate ICL component?}
\label{sec:discussion}

\CX{Discussions of ICL often treat it as a single entity
distinct from the BCG and other galaxies. On this basis a number of authors
have proposed dynamical definitions of an `ICL component' in hydrodynamical
simulations of clusters (for example based on cuts in stellar binding energy)
which isolate roughly the same stars as conventional observational definitions
based on surface brightness \citep[e.g.][]{Murante04, Dolag10, Puchwein10,
Rudick11, Cui14a}. In light of our analysis, it seems likely that these
dynamical criteria are picking out the relatively unrelaxed debris from more
recently accreted progenitors, explaining the differences they obtain in
surface brightness, radius and stellar populations with respect to more relaxed
debris bound to the central potential of the cluster (which they identify with
the BCG).  Defining a separate `ICL' component in this way may be useful in
studying the relaxation state of debris in clusters, if a significant sample of
tracer velocity measurements (e.g.  planetary nebula or globular clusters) can
be observed, and if the formation biases of different tracers can be
understood.}

\CX{However, our results suggest that treating `the ICL' separately from `the
BCG' according to an a priori definition, dynamical or otherwise, is
not a particularly helpful way to interpret photometric observations of
clusters in the context of the $\Lambda$CDM model. Most importantly, the idea
that the ICL has a distinct formation history seems at odds with the firm
theoretical prediction that almost all the stars in BCGs are accreted, and that
these accreted stars have a continuum of binding energies in the dark
matter-dominated potential of the cluster. Hard dynamical separations obscure
the fact that individual progenitors contribute stars over a wide range of
binding energies. Distinguishing components by origin -- in situ or accreted --
is more meaningful from the point of view of $\Lambda$CDM theory, although our
results suggest the in situ contribution is negligible in most massive
clusters. We find no motivation in our results for `fitting and subtracting' an
ICL component in the reduction of BCG photometry.  Instead, starlight in low
surface brightness regions should be consistently accounted for in photometric
quantities measured for individual BCGs, and in their population statistics
such as distributions of size and total mass \citep[e.g.][]{Lin04,Bernardi13}.} 

\section{Conclusions}
\label{sec:conclusions}

We have applied a combination of semi-analytic and $N$-body modelling to the
problem of diffuse light in massive galaxy clusters, following \CB{the
particle tagging methodology introduced by \citet{Cooper10} and used by C13 to
study less massive groups and clusters}. Our approach results in detailed
predictions for the phase space distribution of stars in clusters based on a
standard theory of cluster galaxy formation in $\Lambda$CDM
\citep{Kauffmann93,Springel01,DeLucia04} \CE{constrained by the observed
$z=0$ galaxy stellar mass function (G11). Our $N$-body model has
substantially higher resolution than most hydrodynamical simulations of very
massive clusters}.

Our main results concern the stellar mass surface density profiles of BCGs in
$M_{200}\sim10^{15} \Msol$ clusters and the fraction of stellar mass in these
clusters that is bound to their central potential. They are as follows:

\begin{enumerate}

  \item \textit{BCG mass and morphology are dominated by stars accreted from
    other galaxies, even within \CD{their stellar half-mass radii}.}  The
    suppression of in situ formation in massive haloes (arising in our model
    from the combination of AGN feedback and long radiative cooling
    time-scales) tends to enhance the importance of accretion
    \citep[e.g.][]{Puchwein10}. \textit{In situ} star formation contributes of
    the order of 10 per cent of the total BCG stellar mass and does not
    significantly affect the surface brightness profile beyond 10~kpc at $z=0$.

  \item \textit{Galaxy clusters are rich in dynamical substructure.} In low
    surface brightness regions ($30 \lesssim \mu \lesssim 25 \sbunits$) our
    nine clusters show many faint stellar overdensities with stream and shell
    morphologies. These result from episodic tidal stripping over several
    gigayears of galaxies with stellar masses $\lesssim10$ per cent of that of
    the final BCG. This drives substantial evolution of the BCG SB profile at
    $R\gtrsim100$~kpc between $z=1$ and $0.25$.

  \item \textit{Many cluster galaxies are still being actively stripped at
    $z=0$.} Approximately $20$ per cent of stars accreted by the BCG have been
    stripped from surviving DM subhaloes above our resolution limit.
    Even so, coherent streams with $\mu_{\mathrm{V}} \sim 25 \sbunits$ are
    rare, consistent with counts in very nearby clusters.  Fainter streams are
    more common but have low surface density contrast with respect to other
    diffuse debris. 

  \item \textit{BCG surface brightness profiles have a characteristic double \Sersic{}
    form.} This emerges from the superposition of many separate debris
    components from different progenitors. Taken in rank order of mass, at
    least 40 progenitors are required to account for 90 per cent of the BCG
    mass.

  \item \textit{The profile of each \CX{individual} progenitor debris component can be classified
    (loosely) as either `relaxed' or `unrelaxed'.} `Relaxed' components are
    centrally concentrated and roughly symmetric around the BCG. They are
    associated with early accretion events and/or mergers with low mass ratios
    leading to violent relaxation \citep{White78_mergers_1, Naab06a, Oser10,
    Hilz12}.  `Unrelaxed' components are more diffuse and include the distended
    envelopes of other bright cluster galaxies.

  \item \textit{In circular apertures centred on the BCG, `unrelaxed' debris components
    are characterized by profiles with large effective radius and low \Sersic{}
    index ($n\lesssim2$).} When the total mass in these components is
    significant (as for example in recently merged clusters with several BCG
    candidates) the BCG surface brightness profile breaks to a shallower slope
    at large radii (relative to regions with $\mu \sim 24 \sbunits$, which
    typically have $n \gtrsim 4$). The outer component of a double \Sersic{}
    fit is a good estimate of the nett `unrelaxed' contribution in most cases. 

  \item \textit{Our results support observational evidence for diffuse $n\sim1$
    components in BCG profiles} \citep{Seigar07, Donzelli11}.  Qualitatively,
    such outer exponential components are similar to the `cD envelope'
    phenomenon.  In our simulations, this phenomenon does not occur in all
    clusters of similar $M_{200}$. In some cases, it originates from the tidal
    debris of another massive cluster member, while in others it originates
    from multiple accretion events. A much larger suite of simulations is
    necessary for a statistical study of relationships between the assembly
    history of BCGs and the parameters of their surface brightness profiles.

  \item \textit{The BCG stellar mass fraction in our model has a strong
    $M_{200}$ dependence that extends up to the most massive clusters
    ($M_{200}\sim10^{15} \Msol$)}. This global trend is very similar to that
    seen in observational data. The largest disagreement between our model and
    the data is of the order of $\sim20$~per cent and occurs in the halo mass
    range $10^{13} < M_{200} < 10^{14}\Msol$. Radial trends in the BCG mass
    fraction also agree well.

  \item \textit{\CX{For an SDSS-like observation, ($z\approx0.15$, 1~kpc
    pixels), a surface brightness threshold of $\Sigma \lesssim 10^{6.5}
    \sdunits$ ($\mu_{\mathrm{V}} \sim 26.5 \sbunits$) is a reasonable first
    order cut to isolate the accreted component of the BCG in massive
    clusters.}} In simulated images, this cut recovers 30 per cent of the BCG
    stellar mass, almost all of which is accreted, and requires only a
    $\sim20$~per cent correction for light from other cluster members.

\end{enumerate}

In summary, we find generally good agreement between our model and the
low-redshift galaxy cluster data of \citet{Gonzalez07} and \citet{Donzelli11}.
This implies that the G11 \CD{semi-analytic model assigns plausible}
stellar masses to \CD{at least the most significant} progenitors of present
day BCGs \citep[see also][]{Laporte13b}. The number, mass ratio, timing and
orbits of merger events, which emerge naturally from our $\Lambda$CDM initial
conditions, must also be consistent with the constraints inferred from
idealized merger simulations \citep[e.g.][]{Hilz13}. 

\CE{On the other hand, we find disagreement between our results and the
observational data of \citet{Bildfell08} and a $0.2$~dex overestimate of
$R_{50}$ with respect to an extrapolation of the SerExp relation of
\citet{Bernardi12_arxiv}. At face value, these discrepancies suggest that the
G11 model may overestimate the luminosity of the most massive BCG progenitors
in haloes with $M_{200}\sim10^{15} \Msol$. Further work is required to
understand how uncertainties in the semi-analytic model affect this result. In
particular, the G11 model is known to overpredict the number of galaxies less
massive than the Milky Way at $z \ge 1$ \citep[G11;][]{Henriques:2013aa}. Such
galaxies may contribute to the outer parts of the BCG profile at $z=0$, either
directly or through their contribution to the envelopes of major BCG
progenitors}.

In common with \citet{Contini14}, we find that conclusions regarding diffuse
light in simulations are disproportionately sensitive to the fate of relatively
few massive progenitors (those with $\sim 1$ to $10$ per cent of the BCG
mass\footnote{This contrasts with cluster galaxy luminosity and correlation
functions, which are sensitive to a much larger number of less massive galaxies
\citep[e.g.][]{Kim09, Guo14, Kang14}.}).  Robust quantitative conclusions about
the fraction and profile of ICL require high numerical resolution and well
understood numerical convergence, in terms of both satellite galaxy orbits and
star formation histories. Even at the resolution of Phoenix, the finite
resolution of DM subhaloes introduces uncertainties of up to $\sim20$ per cent.
Modelling the formation and dynamics of galaxies at high enough resolution
across the wide range of scales relevant to galaxy clusters is a challenge to
all models, whether $N$-body, hydrodynamic or semi-analytic. We conclude from
the results above that our hybrid particle tagging approach is successful
enough to merit further comparison with data on the low surface brightness
regions of galaxy clusters.

\section*{Acknowledgements}

We thank the anonymous referee for their helpful comments and suggestions, and
Anthony Gonzalez and Stefano Zibetti for providing their observational data.
This work was carried out as part of the MPA Garching -- NAOC Partner Group of
the Max-Planck Society. Phoenix is a project of the Virgo Consortium.  The
simulations used in this work were carried out on the Lenovo Deepcomp7000
supercomputer at the Super Computing Center of the Chinese Academy of Sciences,
Beijing, and on the Cosmology Machine at the Institute for Computational
Cosmology, Durham.  This work used the DiRAC Data Centric system at Durham
University, operated by the Institute for Computational Cosmology on behalf of
the STFC DiRAC HPC Facility (\url{www.dirac.ac.uk}). This equipment was funded
by BIS National E-infrastructure capital grant ST/K00042X/1, STFC capital grant
ST/H008519/1, and STFC DiRAC Operations grant ST/K003267/1 and Durham
University. DiRAC is part of the National E-Infrastructure. APC acknowledges
the support of a Chinese Academy of Sciences International Research Fellowship
and NSFC grant no.  11350110323. CSF acknowledges support from an ERC Advanced
Investigator grant, COSMIWAY. LG acknowledges support from an NSFC grant (no.
11133003), the  MPG partner Group family, and an STFC Advanced Fellowship, as
well as the hospitality of the Institute for Computational Cosmology at Durham
University. QG acknowledges the support of NSFC grants (nos.  11143005,
11133003) and anNAOC grant (no.  Y434011V01). LG and QG are supported by the
Strategic Priority Research Program `The Emergence of Cosmological Structure'
of the Chinese Academy of Sciences (no.  XDB09000000). VS acknowledges support
by the European Research Council under ERC-StG grant EXAGAL-308037. SW was
supported in part by ERC Advanced Grant 246797 `Galformod'. 

\bibliographystyle{mn2e} 
\bibliography{astro} \bsp
 
\appendix

\section{Numerical resolution}
\label{appendix:numerical}

\begin{figure}
\includegraphics[width=84mm, trim=0.0cm 0cm 0.0cm 0cm, clip=True]{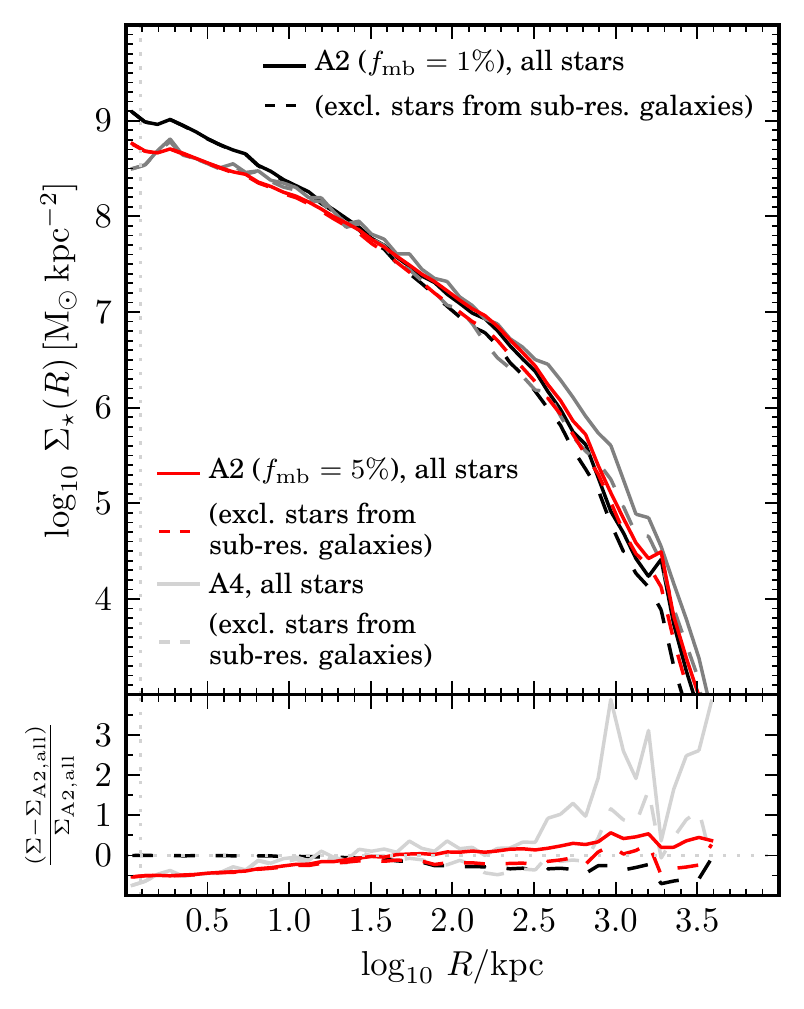}

\caption{Top: the stellar surface density profile of the BCG in Ph-A (black)
\CE{including} stars associated sub-resolution haloes (our default model;
solid) and \CE{excluding} all such stars (dashed). Equivalent profiles are
shown for a larger value of our parameter controlling how deeply new stars are
embedded in their host DM potential, $f_{\mathrm{mb}}=5$~per cent
(red), and for lower resolution (grey). Bottom: residuals around our default
model (black dashed line).} 

\label{fig:restest} 
\end{figure}

\subsection{Convergence}

Fig.~\ref{fig:restest} compares the BCG surface density profile in Ph-A at
`level 2' resolution (black solid line), used throughout this paper, with its
equivalent at the lower `level 4' resolution (a factor of 27 increase in
particle mass; grey solid line). The profile appears well converged overall,
although residuals can be up to  $\sim50$ per cent, with the largest deviations
at $R < 10$~kpc (due to the increased softening length, $\sim 1$~kpc at level
4) and $R \gtrsim 200$~kpc. These differences may be due to stochastic changes
in halo orbits and the timing of accretion events, as well as more rapid tidal
stripping at lower resolution.

\subsection{Tagging fraction}

We use the constant tagging fraction approximation of \citet{Cooper10} and C13,
with $f_{\mathrm{mb}}=1$ per cent. The red solid line in Fig.~\ref{fig:restest}
compares the result for Ph-A with $f_{\mathrm{mb}}=5$ per cent. The effects are
similar to those with lower numerical resolution in the outer part of the halo,
where higher $f_{\mathrm{mb}}$ results in less tightly bound galactic envelopes
that are more easily stripped in the cluster. The behaviour at $R < 10$~kpc is
the result of a larger in situ scale length as described in C13 (in situ stars
dominate this part of the profile in Ph-A). 

\subsection{Sub-resolution haloes}

In the G11 semi-analytic model, galaxies can survive even when their associated
DM halo is lost from the underlying $N$-body simulation after being
stripped below the 20-particle resolution limit of \subfind. \CE{The
time-scale for merging or disrupting these galaxies is determined by simple
semi-analytic recipes} that approximate the orbital evolution of each
satellite.  \CD{This mechanism makes galaxy survival in the semi-analytic
model much less sensitive to the resolution of the underlying $N$-body
simulation} \CE{\citep{Guo14}}.  Galaxies associated with these
sub-resolution haloes are referred to in other work on the G11 model as
`orphans' or `type 2' galaxies.  Comparisons with observed luminosity and
correlation functions constrain the number of galaxies that these recipes need
to `keep alive' to $z=0$ (see figs 14 and 19 of G11). 

In our particle tagging scheme, all galaxies are, by definition, only as
well resolved as their DM haloes. As described in the main text, we
therefore need to decide whether to count stars associated with surviving
sub-resolution semi-analytic galaxies as part of the BCG (following the $N$-body
model) or not (following the semi-analytic model). The `best' (numerically
converged) answer will lie somewhere between the two: the semi-analytic model
does not allow for the tidal stripping of stars from sub-resolution haloes,
while $N$-body particle tagging overestimates it, because some stars may be
bound to the unresolved core and the binding energy of the excess baryons is
ignored. 

Our fiducial choice is to follow the $N$-body simulation and treat all stars
from sub-resolution semi-analytic galaxies as part of the BCG. Our principle
argument in favour of this approach is given in the main text: the halo
resolution limit of Phoenix corresponds \CA{to $\sim0.02$ per cent} of
$M_{200}$ for a Milky Way-mass halo. This extreme mass-loss makes it likely
that a large fraction of the stars in such haloes will have been stripped by the
time the DM is reduced to this limit, regardless of whether the
unresolved core remains bound or not. 

\begin{figure}
  \includegraphics[width=84mm, trim=0.0cm 0cm 0.0cm 0cm, clip=True]{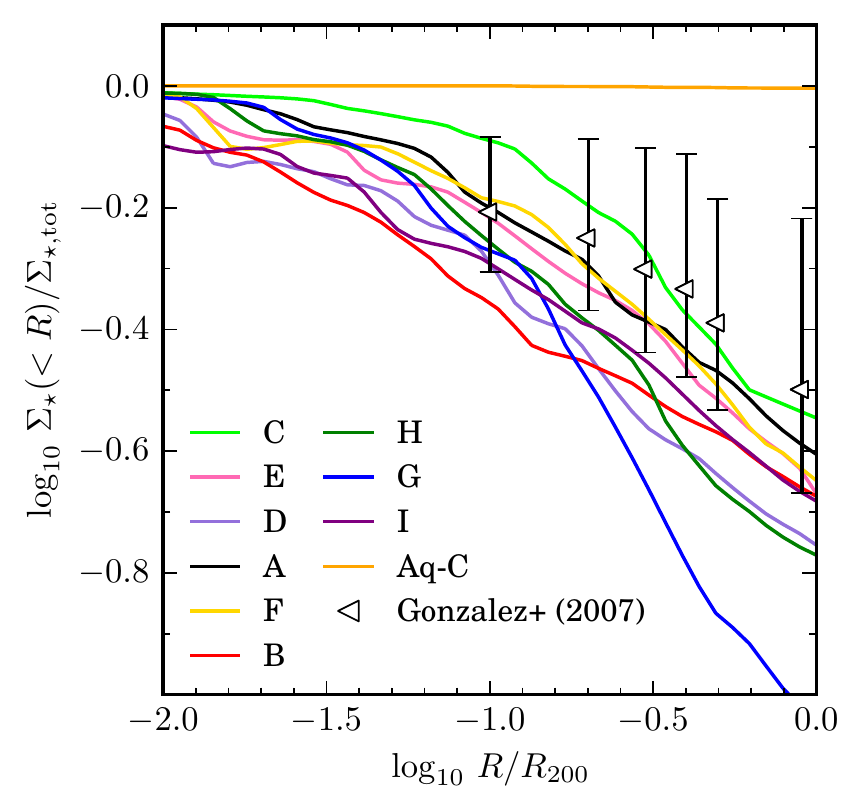}

  \caption{As Fig.~\ref{fig:icl_ratio_radius_cumulative} but excluding all
  stars associated with semi-analytic DM haloes below the resolution
  of the $N$-body simulation at $z=0$ from the definition of the BCG. The
  curves are systematically lower by $\sim0.1$ dex and the variation between
  haloes is greater.}

   \label{fig:icl_ratio_radius_cumulative_XT2}
\end{figure}

\begin{figure}
\includegraphics[width=84mm, trim=0.0cm 0.2cm 0.0cm 0.25cm, clip=True]{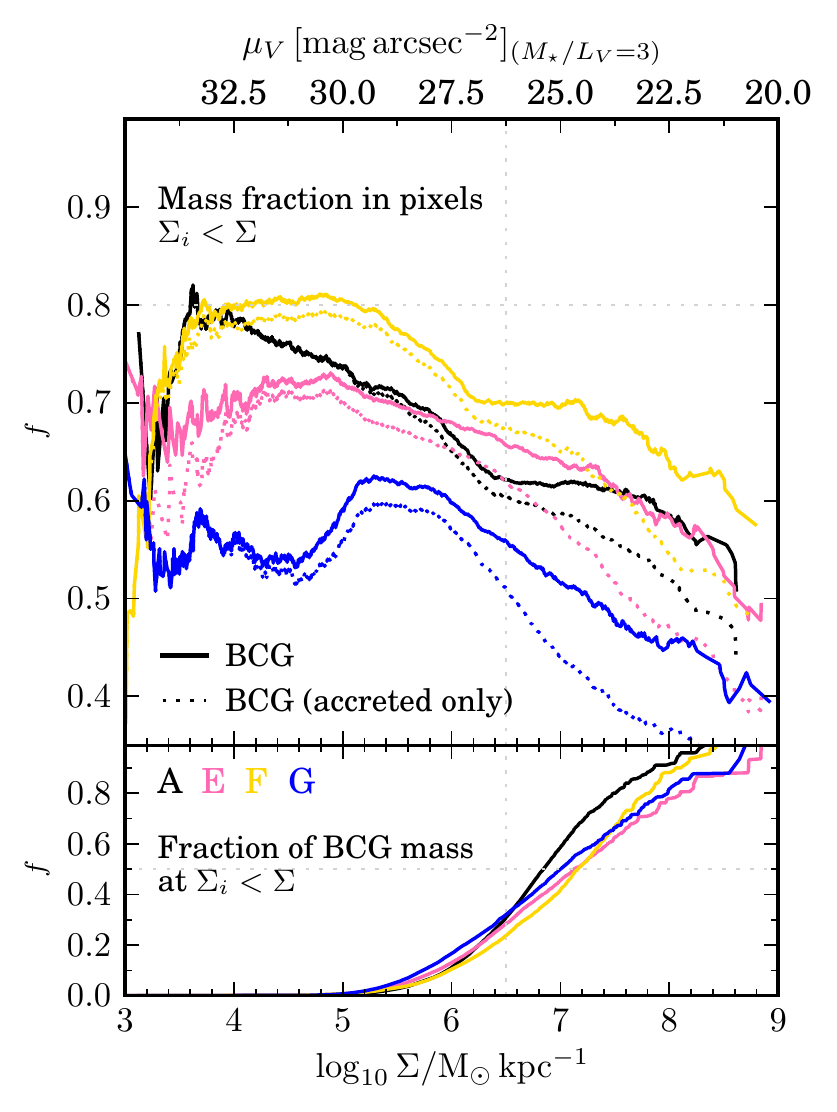}

\caption{As Fig.~\ref{fig:pixel_stats}, but excluding all stars associated with
semi-analytic DM haloes below the resolution of the $N$-body
simulation at $z=0$ from the definition of the BCG.  The relatively large
changes seen in the upper panel are caused by the assignment of stars in very
low surface brightness pixels to cluster galaxies rather than the BCG.
Measurements of this distribution from deep photometry could therefore
constrain models of galaxy disruption in clusters.}

\label{fig:pixel_stats_xt2} 
\end{figure}

This assumption will be less accurate for less massive haloes.  Between 90 and
60 per cent (median $\sim84$ per cent) of the total stellar mass contributed to
the accreted component of the BCG by galaxies associated with sub-resolution
haloes comes from $\lesssim10$ galaxies more massive than $M_{\star}=10^{10}
\Msol$ (i.e.  haloes of $M_{200}\gtrsim10^{12}\Msol$), for which considerable
stellar stripping before the time of halo disruption is likely. Less
massive haloes, e.g. those of $M_{200}\lesssim 10^{10} \Msol$, may retain a
significant fraction of their stars when they cross the resolution limit.
However, these correspond to galaxies of $M_{\star}\lesssim10^{8} \Msol$, which
account for only 1 to 5 per cent of the stellar mass accreted from
sub-resolution haloes.  

\CD{Treating galaxies with sub-resolution haloes as disrupted alters
previously established results from the G11 model to an extent that depends on
$N$-body resolution}. At the relatively high resolution of Millennium II, only
a small fraction of massive galaxies in clusters have sub-resolution haloes and
they do not dominate the agreement between the model and the overall galaxy
population in the field\footnote{It is likely that a slightly larger number of
bound cores will survive in Phoenix compared to Millennium II, because Phoenix
has a similar particle mass and an even smaller force softening length.}. G11
show that 14 per cent of $\sim10^{14} \Msol$ cluster member galaxies more
massive than the Milky Way belong to sub-resolution haloes in Millennium II,
mostly projected within $R< 200$~kpc.  In further support of our fiducial
choice, Figs 14 and 19 of G11 suggest that their model produces more of
these galaxies than observational data imply.  Reducing the number predicted in
Millennium II \CE{by $\sim20$ per cent} would improve agreement with SDSS
results on the radial distribution of galaxies in clusters and the field galaxy
two-point correlation function (Marcel van Daalen, private communication).

Fig.~\ref{fig:restest} shows the effect of stars from sub-resolution haloes on
the overall BCG density profile of Ph-A. The solid black line shows the profile
we adopt in the main text, which includes stars associated with surviving
sub-resolution haloes.  The dashed black line shows the same profile with those
stars excluded. The difference appears small in these logarithmic plots, and
mainly affects radii $R \gtrsim 100$~kpc.  Nevertheless (as shown in
Table~\ref{tab:phoenixstars}), excluding stars from sub-resolution haloes
reduces the total stellar mass of the BCG by $\sim28$ per cent. This fraction
is similar for the other haloes. Fig.~\ref{fig:icl_ratio_radius_cumulative}
shows that this strongly affects the stellar mass fraction attributed to the
BCG beyond $0.1 R_{200}$, and this in turn alters the global 'BCG/total' mass
ratio shown in Fig.~\ref{fig:icl_fraction} by $\sim0.2$~dex.  As expected,
Fig.~\ref{fig:restest} also shows that including stars from sub-resolution
haloes in the BCG definition increases differences in the BCG density profile
due to numerical resolution and the choice of $f_{\mathrm{mb}}$.

\CB{Figs.~\ref{fig:icl_ratio_radius_cumulative_XT2} and
\ref{fig:pixel_stats_xt2} repeat Figs.~\ref{fig:icl_ratio_radius_cumulative}
and \ref{fig:pixel_stats} from the main text. Of all our results, these figures
are most sensitive to the treatment of stars associated with sub-resolution
haloes. The discrepancy between Fig.~\ref{fig:icl_ratio_radius_cumulative_XT2}
and Fig.~\ref{fig:icl_ratio_radius_cumulative}  increases steadily to larger
radii, reflecting the increasing fractional contribution of the uncertain
stellar component further from the BCG seen in Fig.~\ref{fig:restest}. The
small changes in the lower panel of Fig.~\ref{fig:pixel_stats_xt2}, relative to
that of Fig.~\ref{fig:pixel_stats}, confirm that this uncertainty mainly
affects low surface brightness regions, which do not contribute a large
fraction of the total stellar mass of the BCG.}

\label{lastpage}

\end{document}